\documentclass[aip,aps,pop,preprintnumbers,amsmath,amssymb,groupedaddress,floats,nobibnotes,nofootinbib]{revtex4-1}
\usepackage[x11names,dvipsnames]{xcolor} 
\usepackage{graphicx}
\usepackage{amsmath, amsthm, amssymb, mathtools}
\usepackage{mathrsfs}
\usepackage{array}
\usepackage{url}
\usepackage{bm}
\usepackage{hyperref}
\usepackage{accents}

\renewcommand{\div}[1]{\gv{\nabla} \cdot #1} 
\newcommand{\curl}[1]{\gv{\nabla} \times #1} 
\let\baraccent=\= 
\renewcommand{\=}[1]{\stackrel{#1}{=}} 

\newcommand{\jac}{\mathcal{J}}

\newcommand{\ubar}[1]{\text{\b{$#1$}}}
\renewcommand{\vec}[1]{\mathbf{#1}}

\newcommand{\ALMA}{\textbf{ALMA}}

\renewcommand{\curl}[1]{\nabla\times #1} 
\renewcommand{\div}[1]{\nabla\cdot{#1}}
\newcommand{\divp}[1]{\nabla\cdot\left({#1}\right)}
\newcommand{\ddt}[2]{\frac{\partial#1}{\partial#2}}

\newcommand{\antisymc}[1]{\left(\partial_j{#1}+{#1}\partial_j\right)}

\newcommand{\vv}{\vec{v}}

\newcommand{\srho}{\sqrt{\rho}}

\newcommand{\srb}{\sqrt{\ubar{\rho}}}

\newcommand{\mm}{\ubar{\vec{m}}}
\newcommand{\rr}{\ubar{\rho}}
\newcommand{\UU}{\ubar{u}}
\newcommand{\m}{\ubar{m}} 

\newcommand{\gradd}{\nabla^{\dagger}}
\newcommand{\antisymdag}[1]{\left(\gradd\cdot{#1}+{#1}\cdot\nabla\right)}
\allowdisplaybreaks

\newcommand{\fdh}[1]{#1}

\begin{document}
\title{Novel approach to general curvilinear coordinates for plasma fluid applications}
\author{Federico D.~Halpern$^1$}
\email{halpernf@fusion.gat.com}
\thanks{Corresponding author}
\author{Tess N.~Bernard$^1$}
\author{Oleksandr Koshkarov$^2$}
\author{Ronald E.~Waltz$^1$}
\affiliation{$^1$General Atomics, P.O. Box 85608, San Diego, California 92186-5608, United States}
\affiliation{$^2$Los Alamos National Laboratory, Los Alamos, NM 87545, United States}

\date{December 8th, 2025}

\begin{abstract}
In general geometry, plasma fluid equations include nonlinear geometric sources associated with fictitious forces, which pose significant challenges to computer simulations. We reformulate the plasma fluid hierarchy to rigorously preserve geometry and conservation properties critical to numerical simulations, while concealing the geometric sources. In their discrete form, the reformulated models conserve mass, angular momentum, and energy naturally, by simple analogy with the continuum equations. These conservation properties have minimal requirements in discrete space, namely, the anti-symmetry of the first derivative and the orthogonality of the scalar and cross products. By decoupling magnetic geometry, coordinate systems, and numerical discretization, this enables maximum flexibility while preserving physics fidelity. As a testbed, we apply the novel representation to the resistive magnetohydrodynamic system, which involves a complete set of curvilinear operations. We verify the correctness of the approach using steady state liquid metal flows and the classic Orszag-Tang vortex.
\end{abstract}
\maketitle

\section{Introduction}
Magnetized plasmas are governed by a delicate interplay between anisotropy, multiscale dynamics, and geometry, all of which impose stringent constraints on their mathematical and computational treatment. First, the dynamics are aligned to the magnetic field structure, which requires resolving strongly anisotropic behavior \cite{Chew1956, Braginskii1965}. Second, fine-scale dynamics or spatial structures are often embedded in larger, quasi-quiescent regimes, as might be the case in magnetic reconnection phenomena \cite{Biskamp2000, Zweibel2009}. Third, it is important to capture and preserve the natural geometry of a problem, for example, during compression events in inertial fusion devices \cite{Lindl1995, Craxton2015}.

In computer applications, plasma geometry is invoked at distinct, interrelated, yet often conflated levels. At the foundational level lies the model representation, \emph{i.e.} the way in which the equations are written. The standard approaches are the classic Lagrange (advection) and Euler (conservation) representations. Oftentimes, the drift-ordering is used to eliminate fast temporal scales associated with the cyclotron motion\cite{Strauss1976,Pfirsch1996}. Above this level lies the choice of coordinate system. In tokamaks and stellarators, flux or field-aligned coordinates are usually most efficient \cite{Hamada1962, Connor1978, Boozer1981, Beer1995}. A third level involves the integration algorithm, which must be compatible with the chosen representation and coordinates.

Algorithm choices tend to gravitate towards bespoke numerical frameworks, specifically adapted for a certain problem. For instance, a common approach in fluid applications targeting general geometry is to use drift-ordered equations written in advection form, and using field-aligned coordinates\cite{Pfirsch1996,Zeiler1997}. The anisotropy of the plasma flows reduces the problem to quasi-2D geometry, which lends itself naturally to the use of the Arakawa algorithm\cite{Arakawa1966}. Many fluid integrators (\emph{e.g.} \onlinecite{Scott1997,Zeiler1998,Dudson2009}) use this approach. However, this combination requires a slowly evolving background field, weak curvature, and small magnetic fluctuations. Hence, its efficiency comes at the price of generality, since it cannot accommodate strong geometric effects. 

A rigorous treatment of general curvilinear coordinates introduces additional specific challenges. The derivatives of the coordinate basis vectors give rise to the Christoffel symbols. In the Lagrange and Euler formulations, they appear explicitly through the divergence of the kinetic stress, with components
\begin{align}
    \left[\divp{\rho\vv\vv}\right]_i =  \frac{1}{\jac}\partial_j\left(\jac \rho v_i v^j \right) - \Gamma^k_{ij} \rho v^j v_k,
\end{align}where $\jac$ is the Jacobian of the coordinate transformation and $\Gamma^k_{ij}$ are the Christoffel symbols of the second kind. (See Appendix~\ref{AppA_transforms} for a summary of the notation used.) In general geometry there can be up to 18 independent Christoffel symbols, each multiplied by a quadratic non-linear source term. These terms capture fictitious forces which represent generalizations of the centrifugal or Coriolis forces. The viscosity entails additional, even more challenging expressions involving the derivatives of Christoffel symbols themselves.

Although the fictitious forces are crucial for geometry effects and numerical conservation, they can be challenging to implement numerically. To our knowledge, none of the drift-ordered codes retain them -- they are truncated at the \emph{representation} level. Full magnetohydrodynamic (MHD) codes for magnetic fusion often employ cylindrical coordinate systems discretized using finite element methods\cite{Sovinec2004, Czarny2008, Jardin2012}. Some astrophysical and high-energy-physics MHD codes utilize cylindrical or spherical coordinates with explicit treatment of the fictitious forces within a finite volume framework\cite{Fryxell2000, Felker2018}. The most complete theoretical treatment we found\cite{Chacon2004} requires special finite volume numerical techniques to handle the numerical fluxes and an explicit computation of the Christoffel symbols.

This paper deals with plasma geometry primarily at the \emph{representation} level of the model, rather than at the algorithm level. The goal is to \emph{reformulate the fluid equations such that the continuous representation retains the correct geometric effects while avoiding explicit Christoffel symbols} in the nonlinear terms -- even for general, non-orthogonal curvilinear coordinates. Our work is a generalization of the Koopman--von Neumann representation\cite{Koopman1931, VonNeumann1932} of plasma physics\cite{Joseph2020} that has been recently developed and used for plasma simulations\cite{Halpern2018, Halpern2021}. The resulting formulation enables the use of \emph{general curvilinear coordinates without additional geometric source terms}, stemming from the kinetic stress, at the representation level.

It is unsurprising that the equations of motion provided here satisfy the expected conservation properties at the continuous level. The purpose of this work is not to re-establish conservation in continuous space, but to instead formulate the equations in \emph{a form that remains manifestly conserving when represented in a discrete setting}. Although the proofs are presented --for clarity-- using continuous operators, the arguments carry over verbatim to the discrete case unless explicitly stated otherwise. Most proofs rely only on two structural properties: the anti-symmetry of first-derivative operators (\emph{i.e.} $\vec{F} = -\vec{F}^T$ for a continuous or discrete operator $\vec{F}$) and the orthogonality of the inner and cross products\footnote{We remark the intentional avoidance of the Leibniz product rule, which generally does not survive discretization.}. If these conditions are met, discrete conservation follows directly. Consequently, this new framework allows for \emph{structure-preserving discretizations in arbitrary geometry across a broad range of numerical algorithms}. 


The remainder of this paper is organized as follows. In Sec.~\ref{SecII_derivation} we provide a novel representation of the equations of motion, and then derive the resulting conservation mass, momentum, and energy conservation laws. Sec.~\ref{SecIII_validation} describes a minimal consistent discretization and presents several test problems in curvilinear coordinates. Sec.~\ref{SecIV_conclusion} contains some final remarks about this work. For completeness, the appendices present the governing equations written explicitly in anti-symmetry form, rather than in the generalized curvilinear representation used in the main text.

\section{Novel curvilinear formulation of fluids}\label{SecII_derivation}

In order to expose the operator symmetry leading to continuous-discrete analog conservation laws, we evolve volume-weighted variables related to conserved quantities. These are chosen so that the fundamental conserved quantities are quadratic in a Euclidean inner product. To that purpose, we define generalized density, momentum, and internal-energy variables
\begin{align}
\rr \equiv \sqrt{\jac\rho}, \qquad
\mm \equiv \sqrt{\jac\rho}\,\vv, \qquad
\UU \equiv \sqrt{2\,\jac\,u},
\end{align}where $\jac$ is the Jacobian of the coordinate transformation and $u = p/(\gamma-1)$ is the internal energy density of an ideal gas with adiabatic index $\gamma$. The coordinates, denoted by $\boldsymbol{\xi}=(\xi^1,\xi^2,\xi^3)$, are assumed to have a non-singular metric\footnote{\fdh{Coordinate singularities can be handled at the discrete level using standard approaches such as non-singular coordinate patches.}} with a physical differential volume element $dV = \jac\,d^3\xi$. Using these definitions, the conserved mass, kinetic energy, and internal energy take the quadratic forms
\begin{align}
\int \rho\,dV = \int \rr^{\,2}\,d\xi^3, \qquad
\int \frac{1}{2}\rho |\vv|^2\,dV = \int \frac{1}{2}|\mm|^2\,d\xi^3, \qquad
\int u\,dV = \int \frac{1}{2}\UU^{\,2}\,d\xi^3.
\end{align}In Cartesian coordinates ($\jac=1$), the generalized variables reduce to the ones used in earlier formulations\cite{Halpern2020}. The use of Jacobian weighted variables has the effect of absorbing some geometric factors, so that the conservation properties emerge from operator symmetry. The objective behind this choice is to facilitate creating continuous conservation laws with discrete analogs, and, furthermore, it constitutes a generalization of the usual practice of evolving Jacobian-weighted scalars in finite volume discretizations\fdh{\cite{Vinokur1974}}.

To express the equations compactly in these variables, we introduce the Jacobian-weighted divergence operator
\begin{align}
\nabla^\dagger \cdot \vec{A} \equiv \jac \divp{\frac{\vec{A}}{\jac}},\label{eq_div_dagger}
\end{align}which is adjoint to the directional gradient $\vec{A}\cdot\nabla$ under the volume-weighted inner product $\langle f,g\rangle = \int f g \,d\xi^3$,
\begin{align}
\int d\xi^{3} \,f\,(\nabla^\dagger \cdot \vec{A}) \, 
=
- \int d\xi^3 \vec{A}\cdot \nabla f,
\end{align}implying
\begin{align}
\int d\xi^{3} \, f \antisymdag{\vec{A}}f = \int d\xi^{3} f\left(\partial_j A^j + A^j \partial_j\right) f= 0\label{eq_anti_dagger}
\end{align}up to boundary terms. Written in this form, the anti-symmetry operator relevant for conservation resides in the partial derivatives $\partial_j$, minimizing the need to introduce derivatives of the metrics for geometric effects. (By $(\nabla^
\dagger\cdot \vec{A} + \vec{A} \cdot\nabla)f$ we imply the composite operator $\nabla^\dagger \cdot(\vec{A} f) + \vec{A}\cdot\nabla f$.)
. 


The novel representation can be derived starting from the standard fluid equations\cite{Braginskii1965,Lust1959} and is applicable to the entire fluid hierarchy. As a minimal example containing every possible operator, we provide the resistive MHD system. The equations of motion read
\begin{align}
    \partial_t \rr\, +& \frac{1}{2}\antisymdag{\vv}\rr = 0 \label{eq_mass_anti}\\
    \partial_t \mm\,
    +& \frac{1}{2}\left(\vv\gradd + \nabla\vv\right)\cdot\mm + \frac{1}{2}\left(\vv\vv - \vec{I}v^2\right)\cdot\nabla\rr - \vv\times\curl\mm \nonumber \\
    +& \frac{\left(\gamma-1\right)}{\rr}\left(\UU\nabla\UU - \UU^2\vec{G}\right)
    + \jac\frac{\vec{B}}{\mu_0\rr}\times\curl\vec{B}
    + \frac{\jac}{\rr} \nabla \cdot \tau = 0 \label{eq_mom_anti}\\
    \partial_t \UU\, +& \frac{\gamma}{2}\antisymdag{\vv}\UU -\left(\gamma-1\right)\vv\cdot\left(\nabla\UU - \UU\vec{G}\right) \nonumber\\
     +& \frac{\jac}{\UU}\left(\nabla\cdot\vv\cdot\tau - \vv\cdot\nabla\cdot\tau\right) - \frac{\jac}{\UU}\left(\nabla\cdot\vec{B}+\vec{B}\cdot\nabla\right)\times\left(\frac{\eta}{\mu_0} \nabla\times\vec{B} \right)= 0 \label{eq_pres_anti}\\
    \partial_t \vec{B}\, -&\, \nabla\times\left(\vv\times\vec{B}-\frac{\eta}{\mu_0}\nabla\times\vec{B} \right)=0.\label{eq_faraday}
\end{align}Here $\mu_0$ is the magnetic constant in SI units, $\vec{G}=\left(\sqrt{\jac}\nabla\sqrt{\jac}\right)/\jac$ is a geometric factor, and $\eta$ is the plasma resistivity. Starting from the standard formulation, these equations can be obtained from the identities $\jac\nabla\cdot\left(\psi^2\vec{A}\right) = 2\ubar{\psi}\antisymdag{\vec{A}}\ubar{\psi}$, $\jac\nabla\cdot\left(\vec{A}\vec{A}\right) = \ubar{\vec{A}}\left(\nabla^\dagger\cdot\ubar{\vec{A}}\right)+\frac{1}{2}\nabla \ubar{A}^2 -\ubar{\vec{A}}\times\nabla\times\ubar{\vec{A}}$ and other simple manipulations. We remark the unusual formulation of the Ohmic heating term $\left(\nabla\cdot\vec{B}+\vec{B}\cdot\nabla\right)\times\left(\frac{\eta}{\mu_0} \nabla\times\vec{B} \right) = \eta\vec{J}^2$, which ensures energy conservation in the discrete setting\cite{Halpern2020}. \fdh{This expression is strictly equivalent to $\eta\vec{J}^2$ in the continuum, and thus it allows for spatially varying resistivities.} 

The viscosity tensor $\tau$ is rather difficult to evaluate in general curvilinear coordinates even for the simplest case of an ideal unmagnetized Newtonian gas\cite{Holler2014, Oreilly2020}. We have not attempted to reformulate the standard collisional viscous tensor\cite{Braginskii1965}. Instead, we provide a simplified form,
\begin{align}
    \tau & = -\mu \left[-\vec{I} \times \nabla \times \vv + \alpha \vec{I}\left(\nabla\cdot\vv\right)\right],\label{eq_visc_anti}
\end{align}which is useful as a numerical viscosity. \fdh{In the continuum, this formulation is equivalent to the fully-developed curvilinear $\tau$ given in Ref.~\onlinecite{Holler2014}}, but simpler for discrete space implementation and verification. Equation~\ref{eq_visc_anti} is obtained using standard identities for the vector Laplacian, $\nabla\cdot\left(\nabla\vec{A}+\nabla\vec{A}^T\right) = -\nabla\times\nabla\times\vec{A} + 2\nabla\left(\nabla\cdot\vec{A}\right)$ ($\vec{I}$ is the fundamental tensor). Since Eq.~\ref{eq_visc_anti} avoids the covariant derivative, it retains full geometry effects without requiring Christoffel symbols for its evaluation ($\mu$ is a constant viscosity, and $\alpha$ is a numerical constant, usually $4/3$ in a standard Newtonian fluid). Additionally, Eq.~\ref{eq_pres_anti} avoids forming the collisional viscous heating through the contraction $\tau:\nabla \vv$, which also requires derivatives of the metric. Thus, while this representation fails to capture the anisotropy associated with plasma collisions\cite{Braginskii1965}, it provides an accurate accounting of the viscous heating with simple implementation for curvilinear coordinates, \fdh{and is consistent with standard analytical treatments of visco-resistive reconnecting instabilities\cite{Porcelli1987,Cole2006}}.

In the subsections below, the mass, momentum, and energy balance relations are derived, placing emphasis on anti-symmetric or orthogonal forms that can be retained in discrete space by simple analogy. The conservation properties are evaluated up to surface terms. These subsections also serve the purpose of illustrating how the choice of flux operators is related to global conservation.

\subsection{Mass conservation}
The total mass balance is computed starting from Eq.~\ref{eq_mass_anti}, multiplied by $\rr$, and integrated over volume. In component form, and using Eqs.~\ref{eq_div_dagger}~and~\ref{eq_anti_dagger}, we obtain
\begin{align}
   \int dV \ddt{\rho}{t} = -\frac{1}{2}\int d\xi^3\,\rr \antisymc{v^j} \rr = 0,
\end{align}where the integral on the right-hand-side vanishes as long as the first derivative is anti-symmetric. As a comparison, it is instructive to consider a curvilinear formulation based on our previous anti-symmetry approach\cite{Halpern2020}, which in the present notation reads
\begin{align}
    \int dV \ddt{\rho}{t} = -\frac{1}{2} \int \jac d\xi^3 \frac{\rr}{\sqrt{\jac}} \left(\frac{1}{\jac}\partial_j \jac v^j + v^j\partial_j \right)\frac{\rr}{\sqrt{\jac}} = 0.\label{eq_srho}
\end{align}The expression above, which is similar to a numerical strategy used in Ref.~\onlinecite{Morinishi2004}, also conserves mass by anti-symmetry. However, we notice that the flow operator in Eq.~\ref{eq_srho} is no longer anti-symmetric with respect to the evolved quantity $\srho = \rr/\sqrt{\jac}$ -- anti-symmetry applies only as an integral constraint. In addition, the divergence portion of the fluxes contains a $1/\jac$ factor that requires special numerical techniques when the Jacobian vanishes\cite{Morinishi2004}. Instead, we found it convenient to absorb the Jacobian into the conserved quantity, which yields Eq.~\ref{eq_mass_anti}. The flow operator $\partial_j v^j + v^j \partial_j$ is now anti-symmetric with respect to the evolved quantity $\rr$. This strategy constitutes a generalization of the usual practice of evolving Jacobian weighted scalars in finite volume discretizations.

\subsection{Momentum balance}
Momentum is, in general, not conserved in curvilinear coordinates due to the presence of non-conservative, fictitious forces. It is interesting, nonetheless, to examine Eq.~\ref{eq_mom_anti} to illustrate its properties in two limits: (a) Cartesian coordinates and (b) cyclic or ignorable coordinates. Total momentum conservation can be obtained from combining Eqs.~\ref{eq_mass_anti} and \ref{eq_mom_anti}. After simple manipulations we find
\begin{align}
    \partial_t \left(\jac\rho\vv \right)
    +& \frac{1}{2}\left[\left(\vv\cdot\mm\right)\nabla\rr + \rr\nabla\left(\vv\cdot\mm \right)\right]
    + \mm\gradd\cdot\mm
    - \mm\times\curl\mm \nonumber \\
    +& \left(\gamma-1\right)\left(\UU\nabla\UU - \UU^2\vec{G}\right) +
    \jac\vec{B}\times\curl\vec{B}/\mu_0
    + \jac\nabla\cdot\tau
    = 0 \label{eq_mom_anti2}.
\end{align}We first examine the contribution of the Reynolds stress to the momentum balance, followed by the pressure gradient, and the Lorentz force. The viscosity, which is given in divergence form, is assumed to conserve momentum.

The four terms in Eq.~\ref{eq_mom_anti2} involving $\mm$ correspond to a special splitting of the Reynolds stress $\nabla\cdot\left(\rho\vv\vv\right)$. The first two terms contribute to the force balance as some sort of kinetic energy gradient. In curvilinear form, this can be expressed as
\begin{align}
    F^1_i = \int d\xi^3 \frac{1}{2}\left[
    \srb \partial_i \left(\srb v^2\right) +
    \left(\srb v^2\right) \partial_i \srb
    \right]=0,\label{eq_mom_force}
\end{align}which integrates to zero due to the anti-symmetry of the first derivative. The next two terms represent dilation and vorticity mixing. Once fully developed for the $i$'th component, they read as follows:
\begin{align}
    F_i^2= &\int d\xi^3 \left(\mm\gradd\cdot\mm - \mm\times\curl\mm\right)_i\nonumber\\ 
    =& 
    \int d\xi^3 \left[ \m_i \left( \partial_i \m^i + \partial_j \m^j + \partial_k \m^k \right)
    - \m^j\left(\partial_i \m_j - \partial_j \m_i \right) + \m^k\left(\partial_k \m_i - \partial_i \m_k \right)\right].
\end{align}Here $\m_i$, $\m^i$ represent the tangent and reciprocal (or dual) components of $\mm$. Four of the terms above evaluate to zero using anti-symmetry, which leaves
\begin{align}
    F^2_i = \int d\xi^3 \left[ \m_i \left( \partial_i \m^i \right)
    - \m^j\left(\partial_i \m_j \right) 
    - \m^k\left(\partial_i \m_k \right)\right].\label{eq_mom_flux} 
\end{align} 
The next portion of the total momentum corresponds to the pressure force, which yields
\begin{align}
    F^3_i & = \left(\gamma-1\right)\int d\xi^3 \left( \UU \partial_i \UU - \UU^2 G_i \right)\label{eq_mom_pres}.
\end{align}Here we point out if $u=\text{const}$ (constant pressure) Eq.~\ref{eq_mom_pres} vanishes. The $\vec{J}\times\vec{B}$ torque due to the magnetic force is expressed in a manner analogous to the $\mm\times\curl\mm$ term, such that
\begin{align}
    F^4_i = \frac{1}{\mu_0}\int \jac d\xi^3\left[ 
    B^j\left( \partial_i B_j - \partial_j B_i \right) - 
    B^k\left( \partial_k B_i - \partial_i B_k \right)   
    \right].\label{eq_mom_lorentz}
\end{align}Interested readers are referred, for comparison, to a detailed discussion on force balance when a standard formulation using Christoffel symbols is used\cite{Chacon2004}. 
\subsubsection{Cartesian coordinates}
In Cartesian coordinates, the basis vectors are $\vec{x}$, $\vec{y}$, $\vec{z}$, $\jac=1$ and thus $\vec{G}=0$, $\m_i = \m^i = m_i$, $\m_j = \m^j = m_j$, and $\m_k = \m^k = m_k$. (Here the unbarred components denote that factors of $\jac=1$ have been evaluated). Adding Eqs.~\ref{eq_mom_flux}--\ref{eq_mom_lorentz}, the momentum conservation per component reads
\begin{align}
    F^{Cart}_i
    =&\int d\xi^3
    \Bigg\{
    \left[ m_i \left( \partial_i m_i \right)
    - m_j\left(\partial_i m_j \right) 
    - m_k\left(\partial_i m_k \right)\right]
    \nonumber\\&+
    \left(\gamma-1\right) \left( \UU \partial_i \UU \right)
    +\frac{1}{\mu_0}\left[ 
    B_j\left( \partial_i B_j - \partial_j B_i \right) - 
    B_k\left( \partial_k B_i - \partial_i B_k \right)   
    \right] \Bigg\}.\label{eq_mom_cart}
\end{align}We notice that the terms corresponding to the Reynolds stress and the pressure gradient integrate to 0 by anti-symmetry. Next, recalling Eq.~\ref{eq_faraday}, we notice that $\div{\vec{B}}=0$ holds. Adding a factor of $\jac\vec{B}\left(\div{\vec{B}}\right)/\mu_0$ to the remaining terms in Eq.~\ref{eq_mom_cart} gives\footnote{\fdh{Terms proportional to $\vec{B}\left(\div{\vec{B}}\right)$ are sometimes introduced in finite volume discretizations of MHD for numerical stabilization (e.g., \onlinecite{Powell1999}).}}
\begin{align}
    \int d\xi^3 \frac{1}{\mu_0}\left[ 
    B_j\left( \partial_i B_j - \partial_j B_i \right) - 
    B_k\left( \partial_k B_i - \partial_i B_k \right) - B_i \left( \partial_i B_i + \partial_j B_j + \partial_k B_k \right) 
    \right] = 0,
\end{align}with all seven terms integrating to zero by anti-symmetry. Hence, the usual momentum conservation in Cartesian coordinates is recovered. We also remark that, after adding over $x,y,z$ components, the total Lorentz force contribution in Eq.~\ref{eq_mom_cart} cancels out completely, without requiring integration or $\div{\vec{B}}=0$.

\subsubsection{Cyclic coordinates}
It is possible to show that angular momenta are conserved if the coordinate is cyclic. For brevity, we restrict ourselves to a simpler coordinate system with two dimensions $(i,j)$, $dV=\jac d\xi^2$, but the calculation holds in three dimensions as well. Consider a cyclic coordinate such that $\partial_i g^{jk} = 0$, $\partial_i \jac = 0$, $G_i=0$. The angular momentum $\int dV \left(\rho v_i\right)$ is a constant of motion.  Using Eqs.~\ref{eq_mom_flux}--\ref{eq_mom_lorentz}, it is possible to show that
\begin{align}
    F_i^{Cyclic} = \int d\xi^2\Bigg[&
    \m_i \partial_i \left( g^{ii} \m_i + g^{ij} \m_j\right) - 
    \left( g^{ji} \m_i + g^{jj} \m_j\right) \partial_i \m_j + \nonumber\\&
    \left(\gamma-1\right)\UU\partial_i\UU+
    \frac{1}{\mu_0}\left(\jac B^j\right)\left[\partial_i B_j - \partial_j B_i \right]
    \Bigg] = 0.\label{eq_momentum_cons1}
\end{align}The result above is demonstrated as follows. First, terms involving $\m_i \partial_i \m_i$ and $\m_j \partial_i \m_j$ integrate to 0 by anti-symmetry after commuting the metric factors with $\partial_i$. Then, two more terms stemming from the kinetic stress cancel out exactly using $\partial_j g^{jk}=0$. The pressure gradient term is also anti-symmetric. This leaves the contribution of the Lorentz force. Adding a factor of $\frac{1}{\mu_0}\jac\vec{B}\left(\div{\vec{B}}\right)$ to the remaining terms in Eq.~\ref{eq_momentum_cons1} gives
\begin{align}
    \frac{1}{\mu_0} \int d\xi^2 \left\{
    \left(\jac B^j\right)\left[\partial_i B_j - \partial_j B_i \right]
    -B_i\left[\partial_i \left(\jac B^i\right) + \partial_j\left( \jac B^j \right)\right]
    \right\} = 0.
\end{align}The second and the fourth terms cancel using the anti-symmetry of $\partial_j$. The remainder of the terms then resemble Eq.~\ref{eq_mom_flux}, and can be shown to integrate to zero. This can be achieved by writing the expression above in terms of the reciprocal components $B^i,B^j$, commuting the metric factors with $\partial_i$ as needed, and using anti-symmetry.

The simplicity and generality of this proof is remarked, \emph{e.g.} especially in light of the sophisticated procedure needed to recover force balance in Ref.~\onlinecite{Chacon2004}. Eq.~\ref{eq_momentum_cons1} only requires the anti-symmetry of the first derivative, $\partial_i g^{jk} = \partial_i \jac = 0$, and $\div{\vec{B}}=0$. Additionally, mass conservation is required to obtain Eq.~\ref{eq_mom_anti2} from Eq.~\ref{eq_mom_anti}. All of these conditions are feasible in the discrete sense for many numerical strategies.

The momentum conservation properties shown above can be illustrated using polar coordinates as an example. We concentrate on the kinetic stress terms that give rise to fictitious forces. In polar coordinates, we have $g^{rr}=1$, $g^{\theta\theta}=1/r^2$, $\jac=r$. The $\theta$ component of Eq.~\ref{eq_mom_flux} reads 
\begin{align}
    \int dr d\theta \left[\srb v_\theta \partial_\theta \left( \frac{ \srb v_\theta }{r^2} \right) - \srb v_r \partial_\theta \srb v_r \right] = 0,
\end{align}showing that the angular momentum along $\theta$ is conserved. Next, it is instructive to spell out the radial momentum to examine how the fictitious forces appear in the new representation. Naturally, the terms containing the fictitious force are those that involve $\srb v_\theta^2$. After carefully examining Eq.~\ref{eq_mom_anti2}, we select contributions to the radial force balance from both the kinetic force term and the vorticity term. This leads to
\begin{align}
    \frac{1}{2}\srb \partial_r \left(g^{\theta\theta}\srb v_\theta\right) + \frac{1}{2} \left(g^{\theta\theta}\srb v_\theta\right) \fdh{\partial_r} \srb - g^{\theta\theta} \srb v_\theta \partial_r \srb v_\theta = -\frac{\rho v_\theta^2}{r^2},\label{eq_momentum_cent}
\end{align}which corresponds to the centrifugal force. This result is derived using the product rule, and, consequently, it is not an exact result in the discrete sense. As a final step, we examine the limit where $\rho v_\theta=\rho_0\Omega$ is a constant, which leads to
\begin{align}
    \frac{1}{2}\srb \partial_r \left(g^{\theta\theta}\srb v_\theta\right) + \frac{1}{2} \left(g^{\theta\theta}\srb v_\theta\right) 
    \fdh{\partial_r} \srb - g^{\theta\theta} \srb v_\theta \partial_r \srb v_\theta = \nonumber\\
    \frac{\rho\Omega^2}{2}
    \left[ \sqrt{\jac} \partial_r \left(g^{\theta\theta} \sqrt{\jac}\right) - \left(g^{\theta\theta}\sqrt{\jac}\right)\partial_r\sqrt{\jac} \right].
\end{align}The factor involving the metrics is symmetric, negative, and does not vanish under integration. From this exercise, we conclude that the representation of the kinetic stress used in Eqs.~\ref{eq_mom_anti} and~\ref{eq_mom_anti2}) contains the well-known fictitious forces if the coordinate is not ignorable.

\subsection{Energy conservation}
Total energy conservation entails the sum of kinetic ($E_K$), internal ($E_U$), and magnetic ($E_M$) energy contributions that can be computed starting from Eqs.~\ref{eq_mom_anti}--\ref{eq_faraday}
\begin{align}
    E_K = \int dV \partial_t \left(\frac{\rho v^2}{2}\right) =& -\int d\xi^3\,\mm\cdot\Bigg\{
    \frac{1}{2}\left(\vv\gradd + \nabla\vv\right)\cdot\mm 
    + \frac{1}{2}\left(\vv\vv - \vec{I}v^2\right)\cdot\nabla\rr \nonumber\\
    &- \vv\times\curl\mm
    + \frac{\left(\gamma-1\right)}{\rr}\left(\UU\nabla\UU - \UU^2\vec{G}\right)
    + \jac\frac{\vec{B}}{\mu_0\rr}\times\curl\vec{B}
    + \frac{\jac}{\rr} \nabla \cdot \tau
    \Bigg\}\label{eq_energy_kinetic}\\
    E_U = \int dV \partial_t \left(u\right) 
    =& -\int d\xi^3\,\UU\Bigg\{  \frac{\gamma}{2}\antisymdag{\vv}\UU -\left(\gamma-1\right)\vv\cdot\left(\nabla\UU - \UU\vec{G}\right) \nonumber\\
     &+ \frac{\jac}{\UU}\left(\nabla\cdot\vv\cdot\tau - \vv\cdot\nabla\cdot\tau\right)
     -\frac{\jac}{\UU}\left(\nabla\cdot\vec{B} + \vec{B}\cdot\nabla \right)\times\left(\frac{\eta}{\mu_0}\nabla\times\vec{B} \right)
    \Bigg\}\label{eq_energy_internal}\\
    E_B = \int dV \partial_t \left(\frac{B^2}{2\mu_0}\right) =&  \frac{1}{\mu_0}\int \jac d\xi^3\,\vec{B}\cdot\nabla\times\left( \vv\times\vec{B}-\frac{\eta}{\mu_0}\nabla\times\vec{B}\right)\label{eq_energy_magnetic}.
\end{align}Evaluating the total energy $E=E_K+E_U+E_B$ is lengthy, but straightforward. Two of the terms immediately cancel using the orthogonality relations $\mm\cdot\left(\vv\vv-\vec{I}v^2\right)=0$ and $\mm\cdot\vv\times\curl{\mm}$=0, without need for integration. Then, the terms related to the work done by the pressure gradient and the viscosity $\sim \vv\cdot(\nabla p + \div{\tau})$ in Eq.~\ref{eq_energy_kinetic} cancel exactly with their counterparts in Eq.~\ref{eq_energy_internal}. After adding Eqs.~\ref{eq_energy_kinetic}--\ref{eq_energy_magnetic} we obtain
\begin{align}
    E = -\int &d\xi^3\,\Bigg\{
    \frac{1}{2}\mm\cdot\left(\vv\gradd + \nabla\vv\right)\cdot\mm +
    \frac{\gamma}{2}\UU\antisymdag{\vv}\UU + 
    \jac \nabla\cdot\left(\vv\cdot\tau\right)
    \nonumber\\
    &+\frac{\jac}{\mu_0}\vec{B}\cdot\left[\nabla\times\left(\vv\times\vec{B}\right) + \vv\times\nabla\times\vec{B} \right]
    -\jac \divp{\frac{\eta}{\mu_0} \vec{B}\times\nabla\times\vec{B} }
    \Bigg\} = 0.
\end{align}These contributions correspond to the kinetic and internal energy fluxes, the viscous flux, and the Poynting flux. For clarity, we evaluate the components of each term. The kinetic-stress term reads
\begin{align}
    \frac{1}{2}\int d\xi^3 \left[ \left(\m^i v_i\right) \partial_j \m^j + \m^j\partial_j\left( v_i \m^i \right)\right] = 0,
\end{align}and represents the kinetic energy transport $\sim\divp{\rho v^2\vv}/2$. Next, the portion corresponding to internal energy fluxes is written as
\begin{align}
    \frac{\gamma}{2}\int d\xi^3\, \UU\left(\partial_j v^j + v^j \partial_j\right)\UU = 0.
\end{align}The viscosity and the resistive portion of the Poynting flux appear in divergence form, and vanish as well. This leaves the Poynting flux related to the ideal MHD electric field $\vec{E}=-\vv\times\vec{B}$, which can be reduced to
\begin{align}
    \frac{1}{\mu_0}\int d\xi^3
    \left\{ 
    B_i \partial_j \left[\jac\left( v^i B^j - v^j B^i \right) \right] +
    \left[\jac\left( v^i B^j - v^j B^i \right) \right] \partial_j B_i
    \right\} = 0,
\end{align}completing the proof. In conclusion, Eqs.~\ref{eq_mass_anti}--\ref{eq_faraday} satisfy
\begin{align}
    E = \int dV \partial_t \left( \frac{\rho v^2}{2} + u + \frac{B^2}{2\mu_0}\right) = 0,
\end{align}recovered only using the anti-symmetry of the first derivative and the orthogonality of the scalar and cross products.


\section{Implementation, verification, and validation}\label{SecIII_validation}
Numerical verification and validation of the new representation (Eqs.~\ref{eq_mass_anti}--\ref{eq_faraday}) is carried out using the \ALMA{} numerical engine\cite{Halpern2021}. The purpose here is to demonstrate that the MHD equations described above (a) recover known results and (b) that the conservation properties are satisfied in discrete space.

The governing equations are discretized using finite differences on a logically rectangular prism \fdh{(\emph{e.g.} a cuboid}) with a general curvilinear coordinate mapping. To facilitate the implementation of the model equation and its validation, the curvilinear calculus numerical operators needed for the equations were programmed using Python-based symbolic mathematics, and then translated into Fortran code compatible with \ALMA{}. The construction of the operators is based on finite difference methods, as described in our past work\cite{Halpern2020,Halpern2021}, with special consideration to ensure that the symmetry properties are satisfied in discrete space. Flow and force operators are evaluated using centered finite differences. Diffusive, resistive, and viscous terms (the latter two written in curl--curl form) are discretized using diagonally staggered grids\cite{Shashkov1996} to eliminate the discrete null space associated with centered curl operators on collocated grids. 

\fdh{All geometric quantities are computed numerically using the same finite-difference operators used for the right-hand-side of evolution equations. Tangent basis vectors are computed directly from numerical derivatives of the coordinate mapping, while the reciprocal basis vectors and the Jacobian are constructed using Eqs.~25--27 and 31 in Ref.~\onlinecite{Sjogreen2014}. The metric factors $g_{ij}$ and $g^{ij}$ constructed using these basis vectors. Note that, in this context, this procedure does not guarantee exact discrete compatibility with the geometric conservation law. In the present representation, the geometric conservation law would be more intricate than it is in standard conservation form. A full treatment is left for future work.}

\ALMA{} is based on the method of lines, which decouples time and space integration. Time advance is carried out either using the classic Runge-Kutta 4 method (RK4) or implicit midpoint~\cite{Butcher1964}. While the latter method is required to achieve conservation to machine precision, the energy error using RK4 scales like $\Delta t^4$ -- which in practical settings is better than single precision arithmetic.

We remark that the numerical implementation is neither unique nor intended to be optimal. Rather, it serves as a minimal working example of the principal algorithmic components required for a curvilinear MHD formulation.  Alternative discretization choices and coordinate mappings are possible, and they inherit the conservation properties described in the previous sections with minimal requirements. Indeed, some implementations may offer improved efficiency or accuracy for specific applications. \fdh{For instance, the algorithms used here do not enforce monotonicity; however, they reproduce the correct solution in the standard Sod shock tube test (see Appendix~\ref{AppD_Sod}).} Two representative examples of our minimal approach are presented below.

\subsection{Steady-state MHD flows}
We consider the steady-state flow of an electrically conducting fluid in a rectangular channel surrounded by rigid walls. This problem has a known analytical solution\cite{Shercliff1953} and benchmark flow rates\cite{Smolentsev2015} that can be used to verify the correct implementation of \fdh{Eqs.~\ref{eq_mom_anti} and~\ref{eq_faraday}.} 

The problem setup can be described as follows. A uniform magnetic field $\vec{B}_0=B_{0}\vec{y}$ is imposed parallel to the vertical walls, and the flow is driven by a constant pressure gradient. The velocity satisfies no--slip boundary conditions, and the magnetic field satisfies perfectly conducting boundary conditions
\begin{align}
\vv\big|_{\text{wall}} &= 0, &
(B_x,B_y,B_z)\big|_{\text{wall}} &= (0,B_0,0).
\end{align}The normal derivatives of the tangential magnetic field components are set to zero.

The induced magnetic field is usually negligible compared to the imposed field, which justifies the use of the quasi-static MHD limit in classical treatments\cite{Shercliff1953, Smolentsev2015}. The Lorentz force, arising from currents induced by the fluid motion across $\vec{B}_0$, strongly damps the bulk flow. This results in slow velocity cores surrounded by steep boundary layers. The key parameter characterizing the solutions is the Hartmann number,
\begin{align}
Ha = B_0 L \sqrt{\frac{1}{\mu \rho \eta}},
\end{align}which measures the relative strength of Lorentz and viscous forces in the flow. As $Ha$ increases, velocity gradients transverse to the magnetic field are increasingly suppressed in the bulk, and the required shear is confined to thin boundary layers adjacent to the walls. For Shercliff flows, the shear layers normal to the magnetic field have a characteristic width scale $\delta \sim L/Ha$. Access to solutions at very large Hartmann number is important for modeling liquid metal breeder blanket flows, such as LiPb, where strong magnetic fields and high electrical conductivity lead to $Ha > 10^4$. For this reason, this problem has been adopted as a standard benchmark for liquid metal MHD flows\cite{Smolentsev2015}.

Unlike most classical treatments of Shercliff flow, which assume the quasi-static MHD limit (magnetic Reynolds number $Rm \ll 1$), the present simulations are fully inductive by evolving the magnetic field self-consistently using Faraday's law -- Eq.~\ref{eq_faraday}. The pressure is held fixed and thermal effects are neglected, while the density is taken to be constant in order to mock up incompressibility within the compressible formulation used by \ALMA{}. \fdh{Consequently, this problem  only tests Eqs.~\ref{eq_mom_anti} and~\ref{eq_faraday}.} The imposed background magnetic field is sustained through the boundary conditions, and the flow is driven by a constant gravity term. 

\begin{figure}
    \centering
    \includegraphics[width=0.32\linewidth]{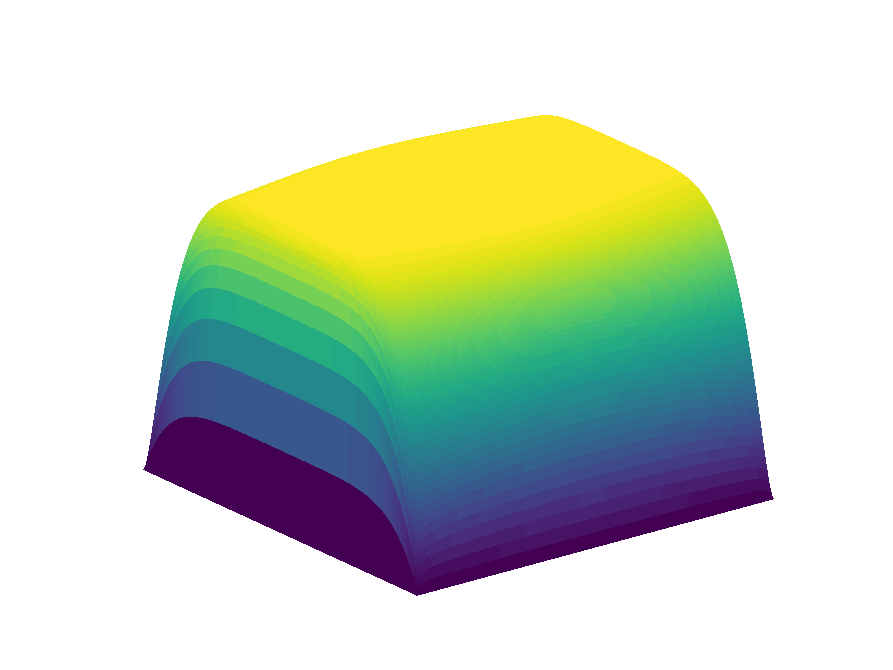}
    \includegraphics[width=0.32\linewidth]{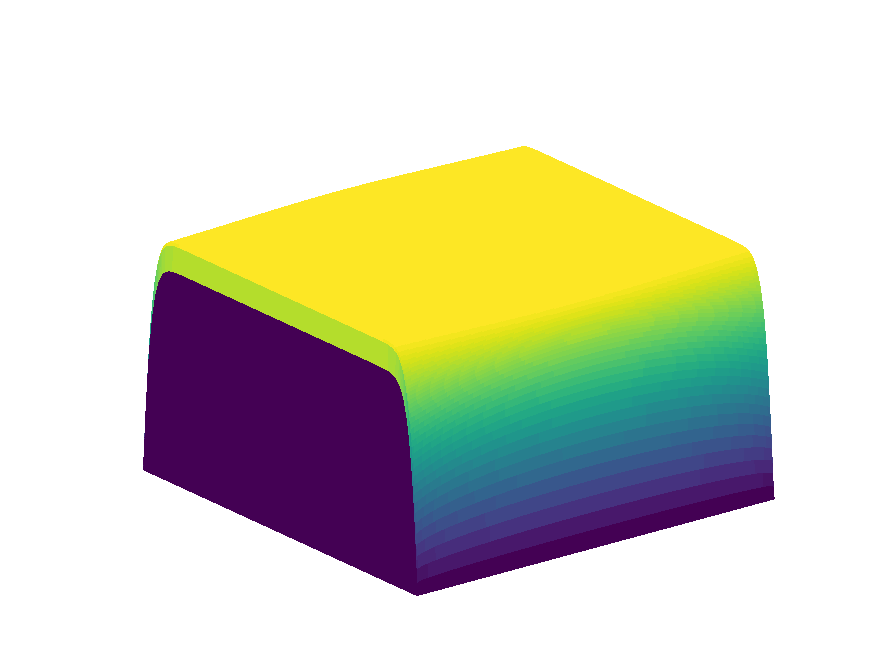}
    \includegraphics[width=0.32\linewidth]{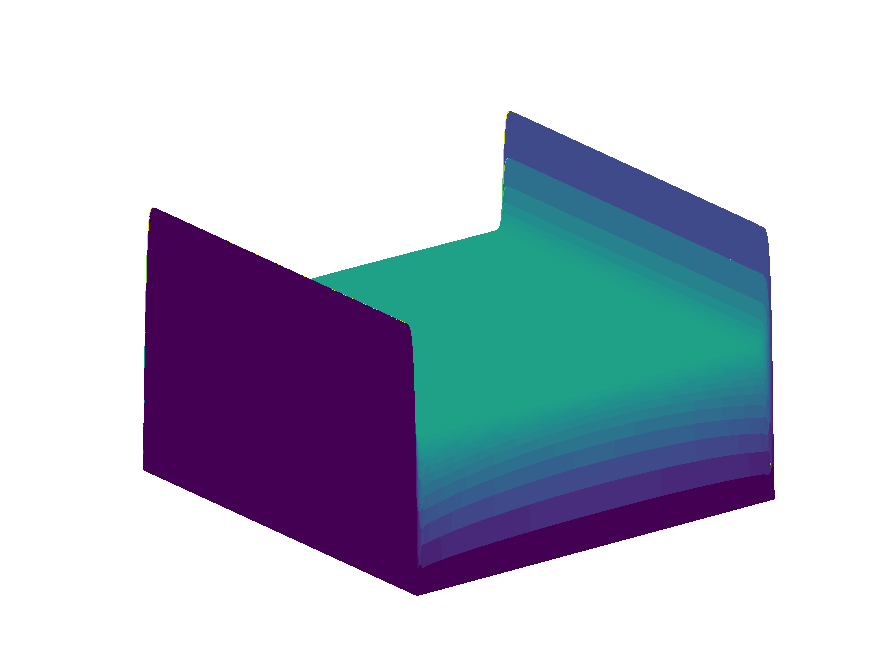} 
    \caption{Axial flow profiles obtained for Shercliff flows at $Ha=50$ (left), $Ha=500$ (center) and $Ha=5000$ (right) at $(N_x,N_y)=320$.}
    \label{fig_Shercliff_profiles}
\end{figure}

\begin{figure}
    \centering
    \includegraphics[width=0.45\linewidth]{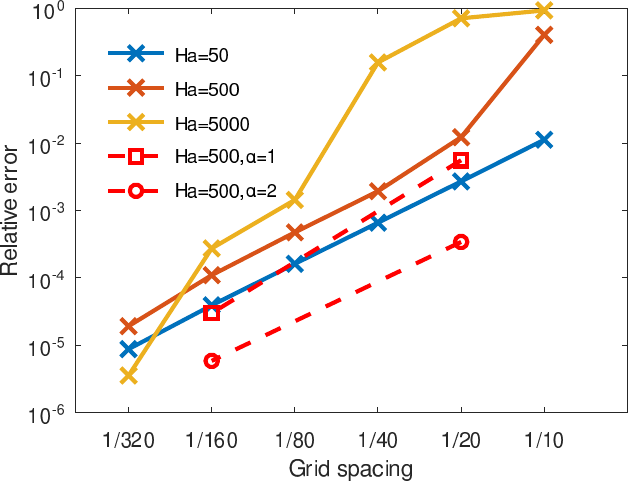}
    \caption{Relative error $\epsilon = (Q-Q_{ref})/Q_{ref}$ as a function of the grid spacing for Shercliff flows with $Ha=50,500,5000$. $Q=\int v_z dV$ is the integrated axial flow rate. \fdh{The red squares ($\alpha=1$) and the red circles ($\alpha=2$)} indicate simulations carried out using a smooth curvilinear transform (Eq.~\ref{eq_transform_shercliff}) that improves accuracy near the walls.}
    \label{fig_Shercliff_error}
\end{figure}

We carry out simulations at $Ha=\{50,500,5000\}$ with grid sizes $(N_x,N_y)=(20,20)$ up to $(640,640)$ inside a square domain $-L \leq x,y \leq +L$ with $L=1$. The simulations are setup with $\eta=\mu=\rho=1$, and the imposed magnetic field is scaled to change the Hartmann number. As an example, Fig.~\ref{fig_Shercliff_profiles} shows the obtained flow profiles at $Ha=50$ (left), $Ha=500$ (center), and $Ha=5000$ (right) \fdh{at $(N_x,N_y)=320$}. The flows are normalized to their maximum values. The steeper walls are the Hartmann layers, while the walls with shallower gradients are parallel to $B_y$ and have steepness $\delta\sim L/\sqrt{Ha}$. The $Ha=5000$ solution presents overshoots which likely stem from the lack of monotonicity enforcing algorithms. Figure~\ref{fig_Shercliff_error} shows the relative error $\epsilon = (Q-Q_{ref})/Q_{ref}$ in the integrated flow rate $Q = \int v_z dV$. The reference flow is obtained from numerical quadrature of the analytical solution\cite{Shercliff1953}, and matches Ref.~\onlinecite{Smolentsev2015}. We find that the error decreases with the square of the grid spacing, although we point out some exceptions. First, at $Ha=5000$ and large grid spacing, the flow is severely underresolved, and the error approaches $\mathcal{O}(1)$. This is due to the width of the Hartmann layer and Shercliff layers, $\delta/L \sim Ha^{-1}\approx 10^{-4}$ and $\delta/L \sim Ha^{-1/2}\approx 10^{-2}$ respectively. Then, at the highest resolution --for unknown reasons-- the integrated error suddenly dips. We also note overshoots on the flow profiles at $Ha=5000$, since the algorithm does not guarantee flow monotonicity.

\begin{figure}
    \centering
    \includegraphics[width=0.32\linewidth]{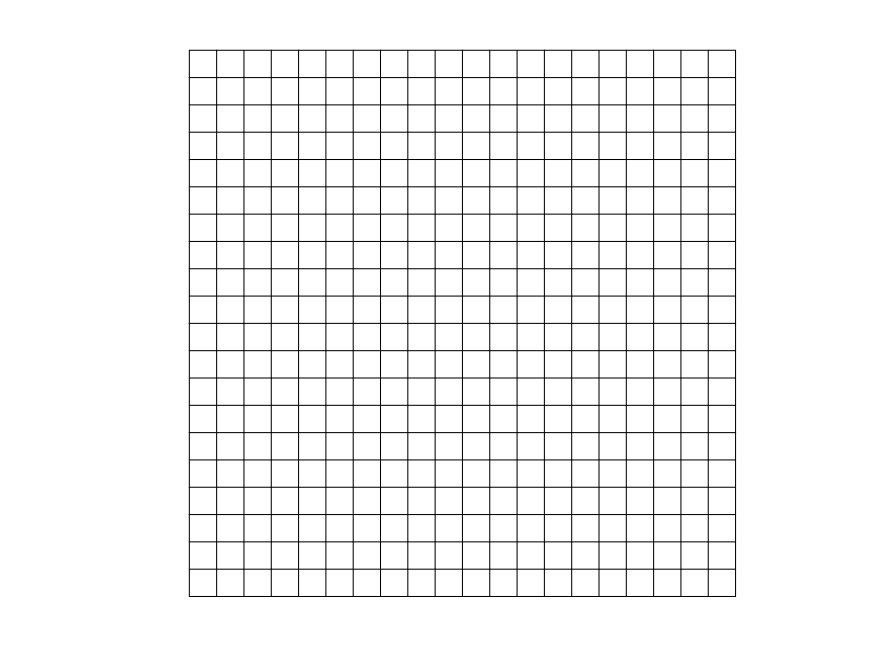}
    \includegraphics[width=0.32\linewidth]{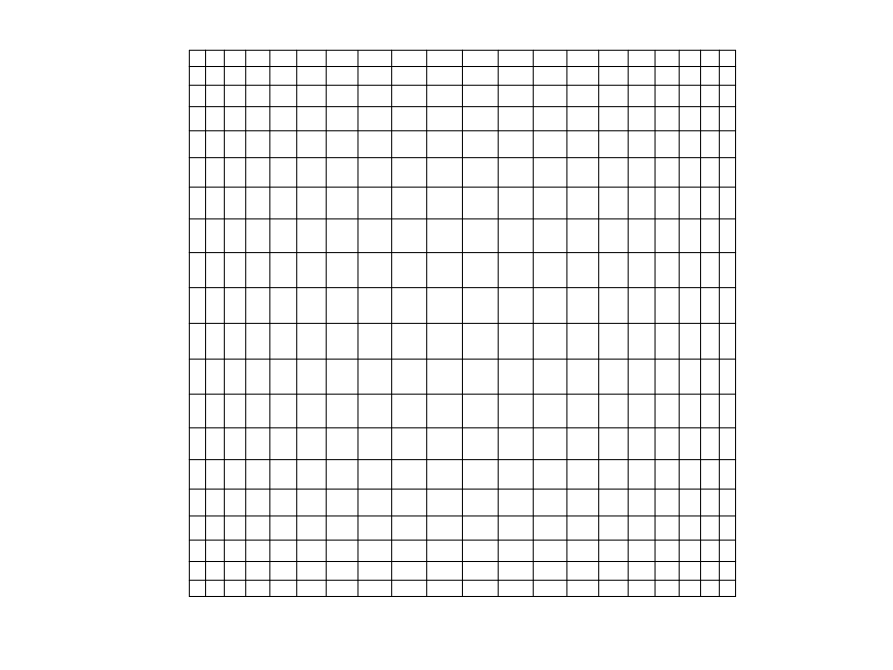}
    \includegraphics[width=0.32\linewidth]{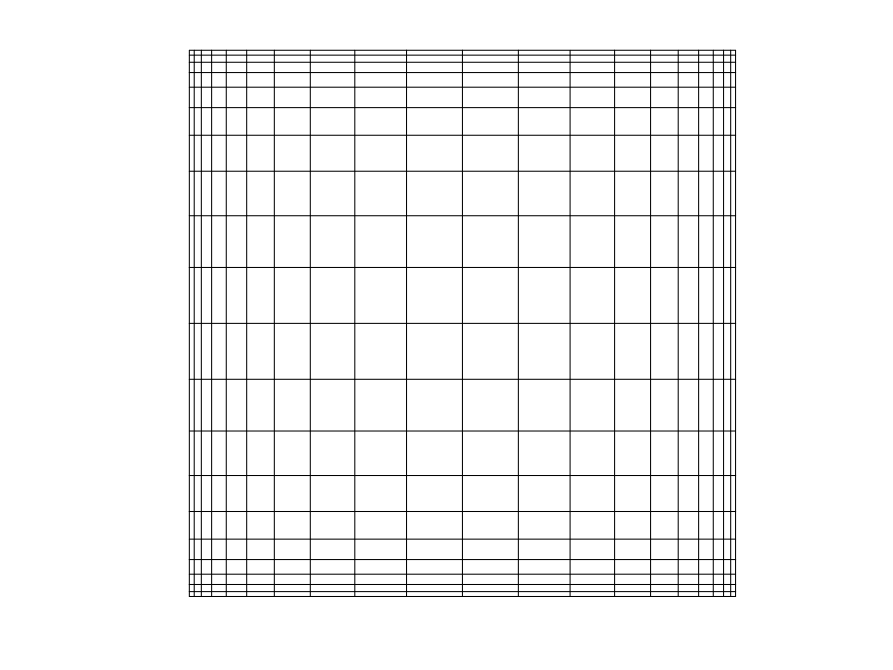}
    \caption{Physical $x,y$ coordinates resulting from Eq.~\ref{eq_transform_shercliff} with $S=0,1,2$ (left, center, right) and $\chi_m$ set to the center of the domain. }
    \label{fig_Shercliff_grids}
\end{figure}
The dominant discretization error at large Hartmann number originates in the unresolved Hartmann and Shercliff boundary layers. Thus, increasing the resolution near the walls can reduce the global error without increasing the total number of grid points. As an initial test for our general coordinate capabilities, we have implemented the following coordinate transform:
\begin{align}
x(\chi)=& \chi_0 + \frac{1}{2}L_x\left\{1 + \frac{\tanh\left[\alpha_x\left(1-2\frac{\chi-\chi_m}{L_x} \right) \right]}{\tanh\alpha_x} \right\},\label{eq_transform_shercliff}
\end{align}with an analog expression for the $y$ coordinate. This transformation has the effect of significantly increasing the resolution near the boundary layers using smooth curvilinear compression. \fdh{Simulations with $\alpha_x = \alpha_y = 1,2$, $\chi_m=0$, $Ha=500$, $(N_x,N_y)=(320,320),(40,40)$ were carried out, and marked in Fig.~\ref{fig_Shercliff_error} as red squares and red circles.} They show an order of magnitude improvement at fixed resolution. The improved accuracy is obtained using a uniform computational grid at a fixed number of degrees of freedom. \fdh{As an illustration, Fig.~\ref{fig_Shercliff_grids} shows the $x,y$ coordinates generated by Eq.~\ref{eq_transform_shercliff} using the simulation parameters.}

This problem provides a stringent test of the correct implementation of both the resistive and viscous stress operators $~\sim\nabla\times\nabla\times$ in curvilinear coordinates. The steep shear layers are sensitive to errors in second-order differential operators and metric terms, particularly near the walls. The asymptotic 2nd order error convergence and the error reduction observed using a coordinate transform (Eq.~\ref{eq_transform_shercliff}) provide direct verification of the consistency and accuracy of these operators within a fully inductive framework. Another benchmark, now showcasing the ability to conserve energy in arbitrary curvilinear coordinates, is shown below.
\subsection{Energy conservation using non-orthogonal curvilinear coordinates}
\begin{figure}
    \centering
    \includegraphics[width=0.4\linewidth]{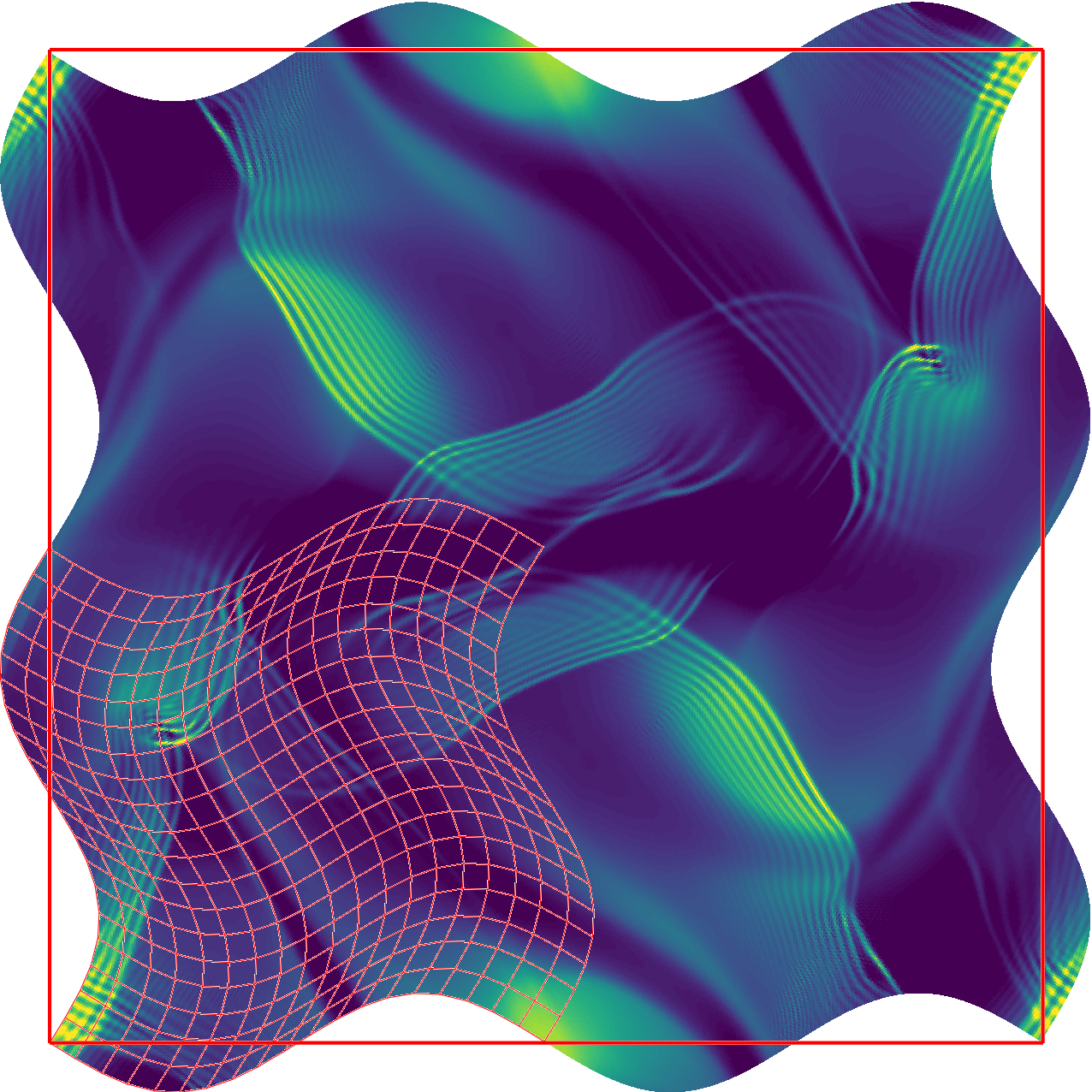}
    \includegraphics[width=0.4\linewidth]{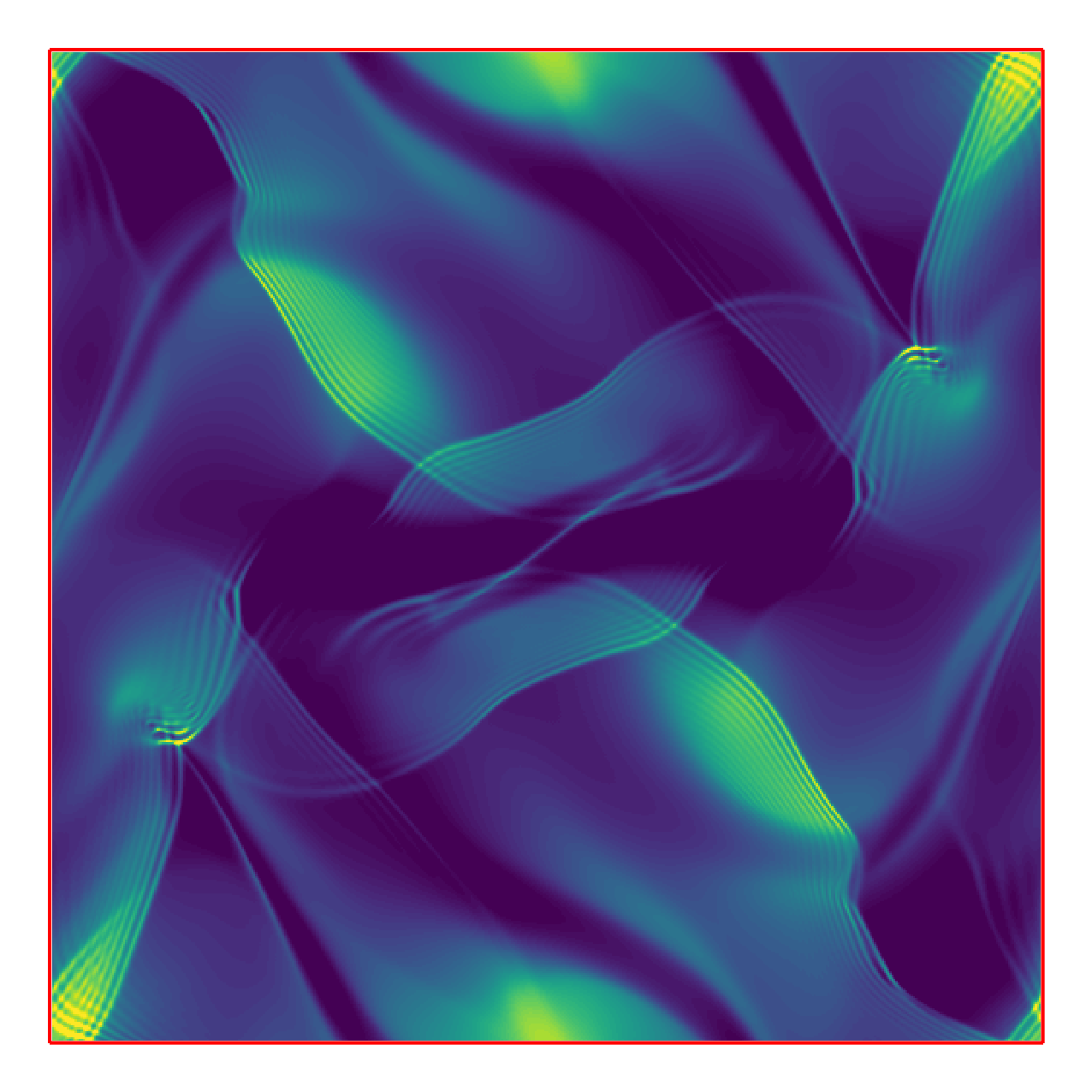}\\
    \includegraphics[width=0.4\linewidth]{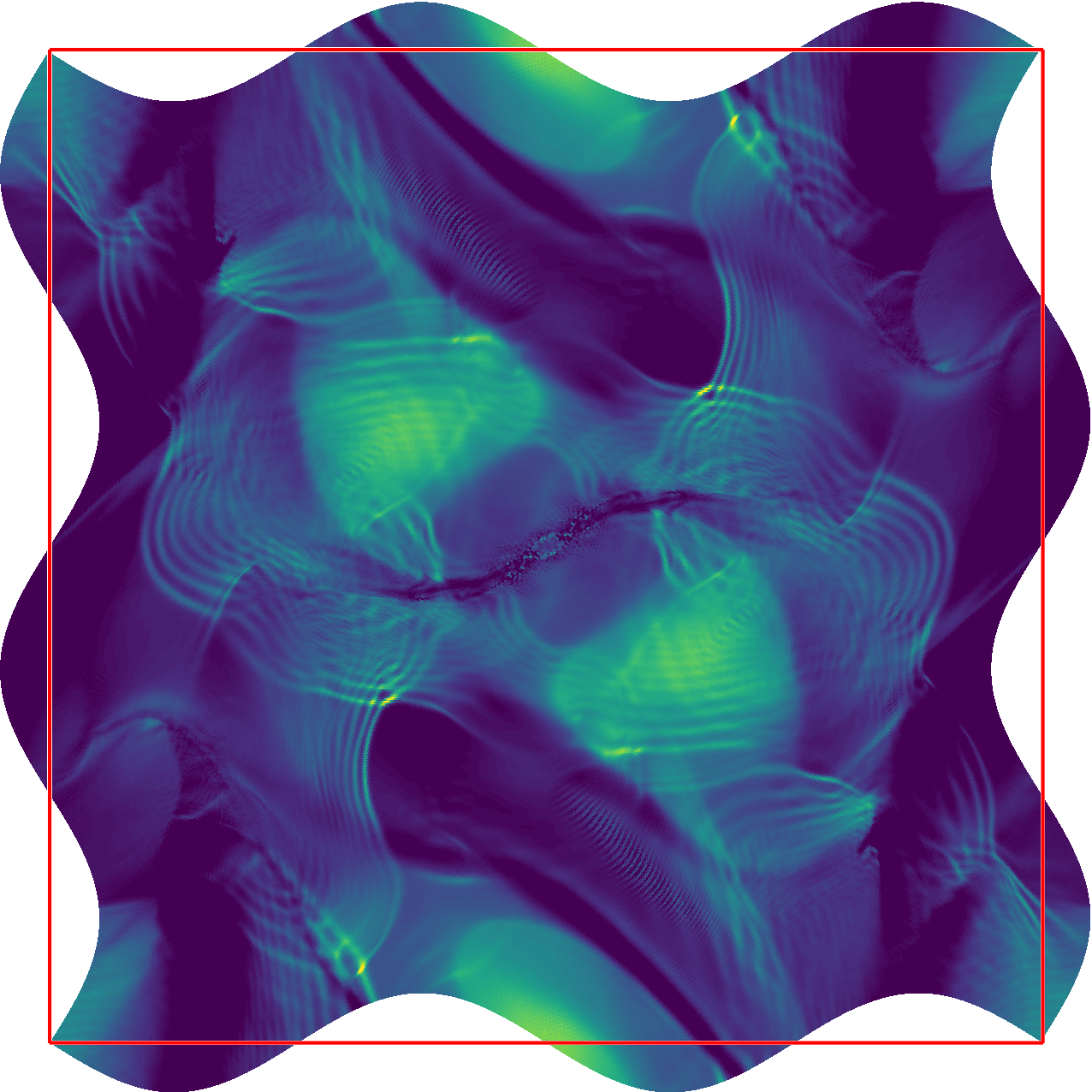}
    \includegraphics[width=0.4\linewidth]{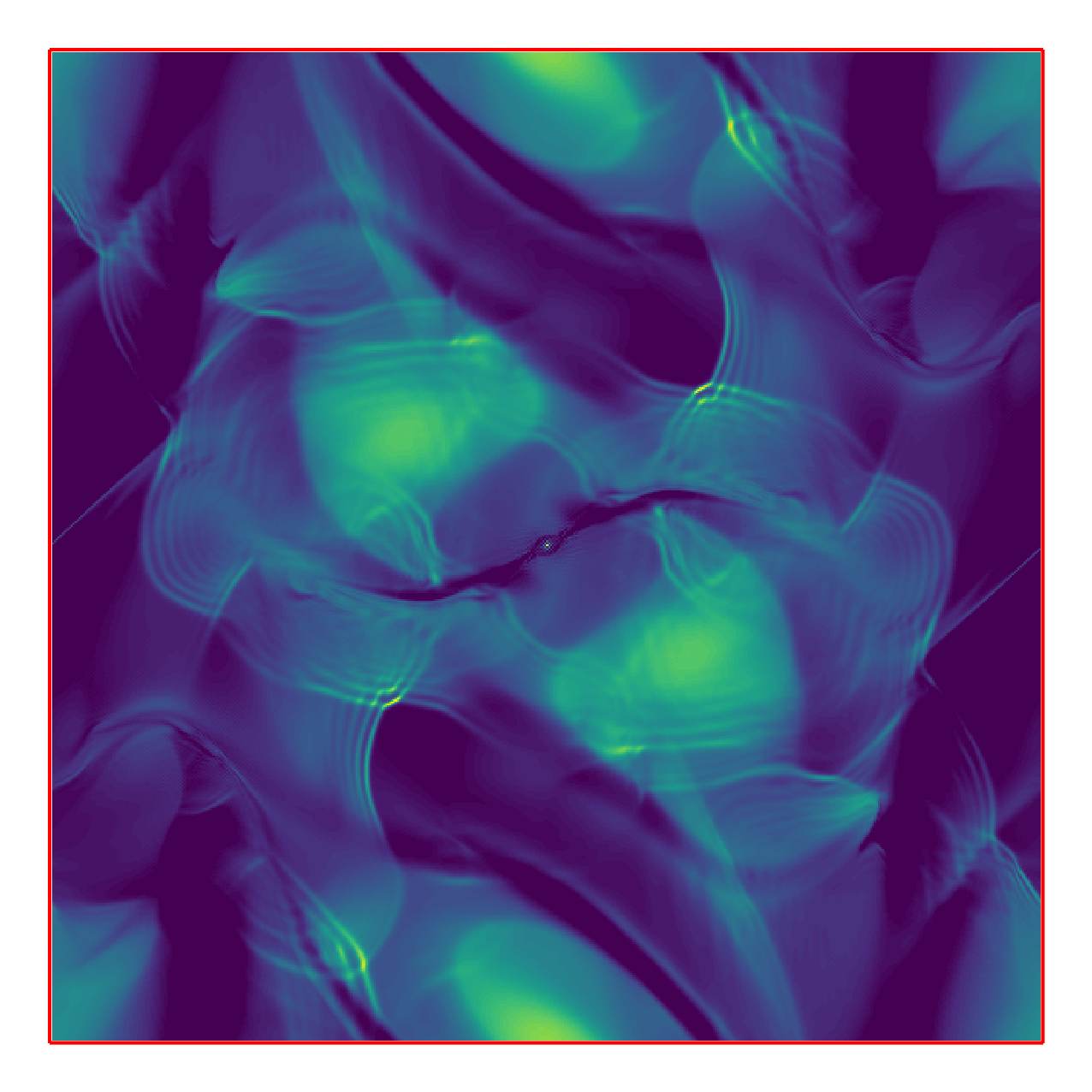}\\
    \includegraphics[width=0.4\linewidth]{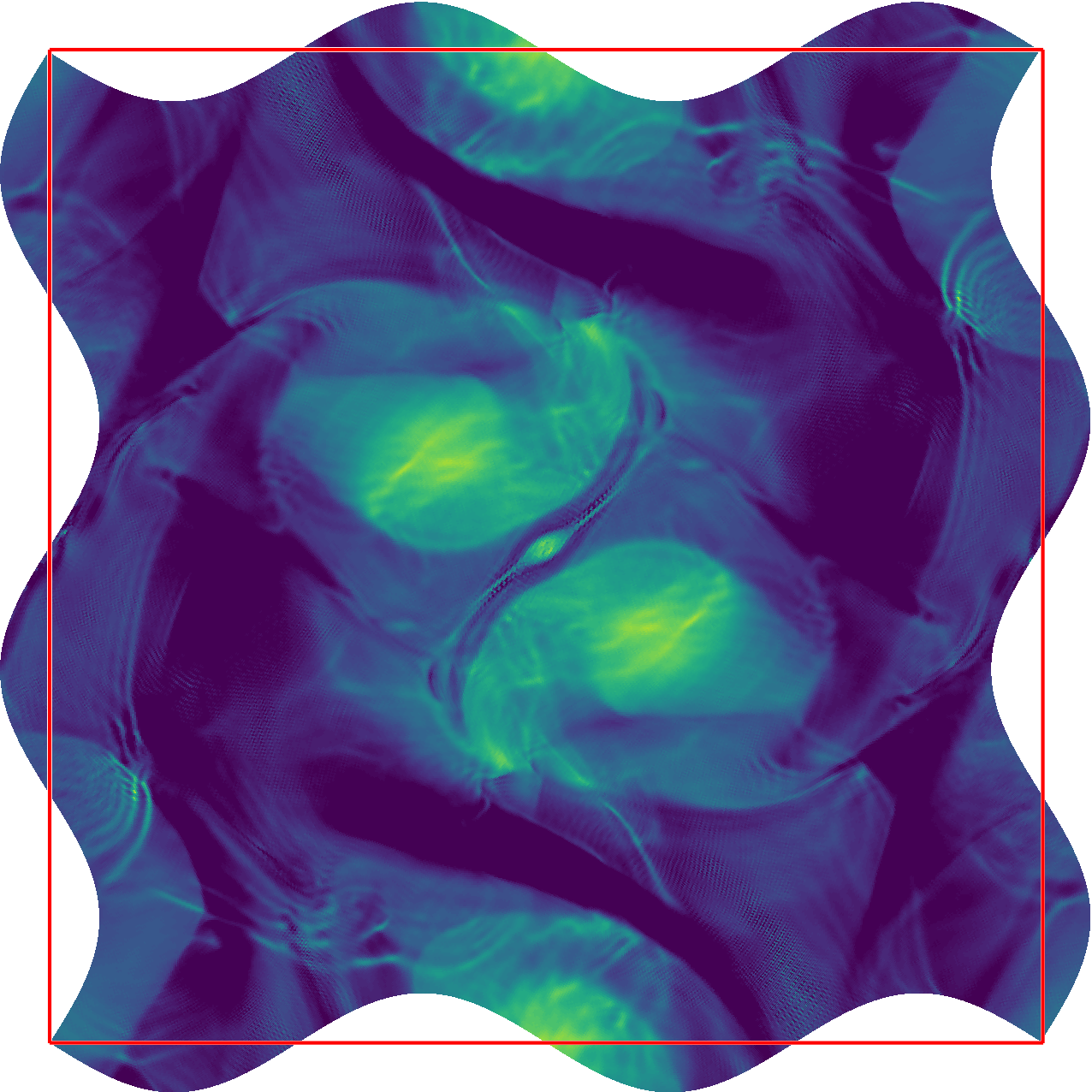}
    \includegraphics[width=0.4\linewidth]{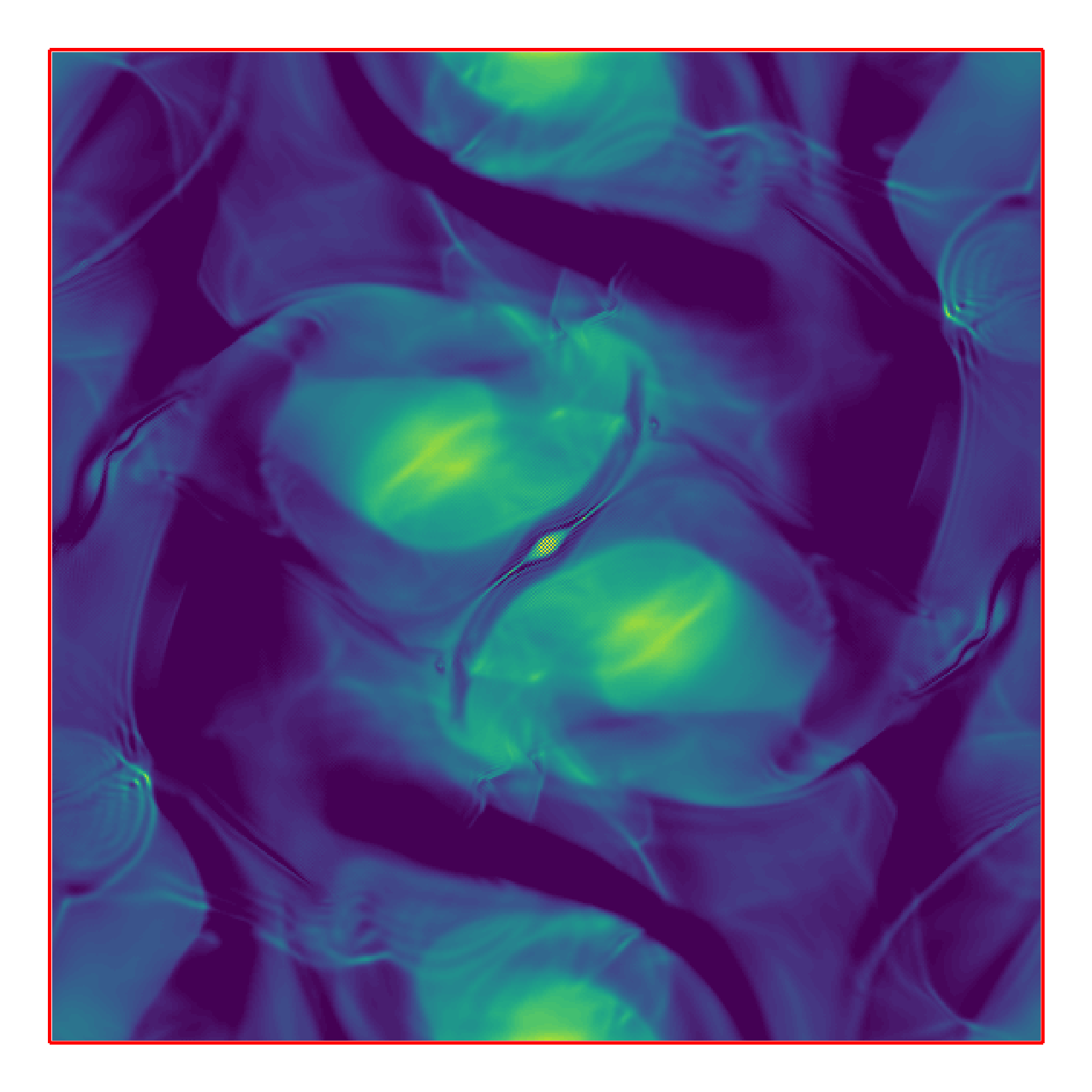}
    \caption{Plasma density at $t=0.25,~0.375,~0.5\tau_A$ (top, center, bottom) for simulations of the Orszag-Tang vortex problem using either curvilinear (Eq.~\ref{eq_coord_OT}) or Cartesian coordinates (left, right). The unit square is marked as a red line.}
    \label{fig_OT_comparison}
\end{figure}
The Orszag--Tang vortex is a standard two-dimensional MHD benchmark originally introduced to study current sheet formation and the transition to turbulence \cite{Orszag1979}, and is widely used to assess nonlinear wave interactions and control of numerical magnetic monopoles (e.g.~\onlinecite{Fryxell2000,Dudson2009,Felker2018}). Since the problem statement involves a doubly periodic unit square Cartesian domain, it can be adapted to demonstrate energy conservation in curvilinear non-orthogonal coordinates. To that purpose, we employ the following coordinate transform
\begin{align}
    x(\chi,\zeta) & = \chi  + a\sin\left(k_\zeta\,\zeta\right)\\
    y(\chi,\zeta) & = \zeta + b\sin\left(k_\chi\,\chi\right),\label{eq_coord_OT}
\end{align}which leads to a non-orthogonal metric $g^{ij}$ and furthermore $\jac=f(\chi,\zeta)$, \emph{i.e.} no ignorable coordinates. The initial conditions (in Cartesian coordinates) are given by
\begin{align}
\rho(x,y,0) &=\frac{\gamma^2}{4\pi}, \qquad p(x,y,0)=\frac{\gamma}{4\pi},\nonumber\\
\vec{v}(x,y,0) &= -\sin(2\pi y)\,\vec{x} + \sin(2\pi x)\,\vec{y}, \nonumber\\
\vec{B}(x,y,0) &= -B_0\sin(2\pi y)\,\vec{x} + B_0\sin(4\pi x)\,\vec{y},\nonumber
\end{align}with $\gamma=5/3$ and $B_0=1/\sqrt{4\pi}$. This choice yields $T(0)=1/\gamma$, $c_s=1$, $v_a=1/\gamma$, and $\beta=2\gamma$, matching the setup used in Refs.~\onlinecite{Fryxell2000,Dudson2009,Felker2018,Halpern2021}. The velocity and magnetic fields are divergence-free at $t=0$, and using our algorithms, the magnetic field remains divergence free thereafter.

\begin{figure}
    \includegraphics[width=0.4\linewidth]{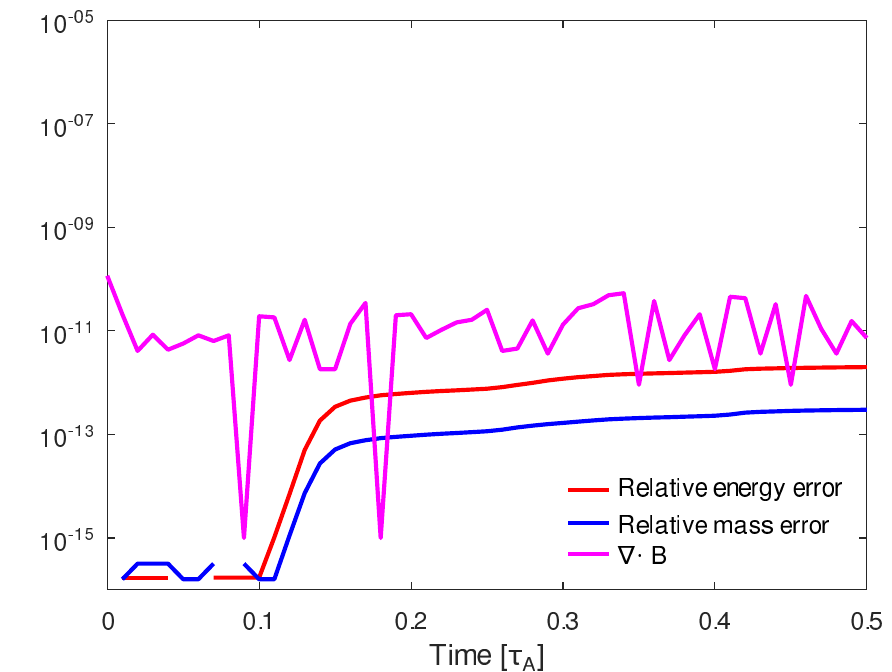}
    \caption{Time traces of total mass, total energy and $\nabla\cdot\vec{B}$ for simulations of the Orszag-Tang vortex problem using curvilinear coordinates. Time integration is carried out using the implicit midpoint method.}
    \label{fig_Time_Traces}
\end{figure}

The discretization described earlier in this section is employed, using second-order centered finite differences in space and an implicit midpoint time integrator. The coordinate transform introduces a moderate amount of distortion, \fdh{with $a=b=0.05$, $k_\zeta=k_\chi=2$}. The domain evolved corresponds to the (curvilinear) periodic unit square $0 \le \chi,\zeta < 1$ Aside from small artificial dissipation terms, no explicit shock-capturing or monotonicity-enforcing mechanisms are applied. Despite initially uniform thermodynamic fields, steep gradients and interacting shocks form rapidly, concentrating near the domain center, which acts as a stagnation point. Figure~\ref{fig_OT_comparison} shows simulation results \fdh{with resolution $(N_\chi,N_\zeta)=(512,512)$} compared side-by-side with simulations using Cartesian coordinates ($a=b=0$). We plot the density at early, intermediate, and late times, $t=0.25, 375, 0.5$, to demonstrate the qualitative nonlinear behavior. \fdh{On the lower-left quadrant of one figure, we overlay a mesh generated using the distorted coordinates.} Despite the geometric distortion associated with the curvilinear transform, the qualitative dynamics and dominant structures are rather similar. 
\fdh{To check for consistency, we computed center-of-mass of the curvilinear simulations, which show a small drift $\sim 10^{-6}$ away from the center of the domain. This phenomenon could be attributed to the incomplete discrete geometric compatibility of the metric terms with the differential operators, or to the lack of momentum conservation in the distorted coordinates.} Finally, using time traces of the volume integrals of the kinetic, magnetic, and internal energy, it is confirmed that the total energy is conserved to 12 digits of accuracy, while the magnetic field remains divergenceless to about 11 digits (Figure~\ref{fig_Time_Traces}). 

Altogether, it is noteworthy that the curvilinear simulation remains equivalent to the Cartesian case and preserves exact conservation, despite being evolved in a coordinate system that is not aligned with the natural symmetries of the problem.

\section{Conclusion}\label{SecIV_conclusion}
The main results of this paper is a novel representation of the fluid equations in general curvilinear geometry such that explicit \fdh{Christoffel} symbols do not appear. We find a new form that remains manifestly conserving when represented in a discrete setting. This enables the use of general curvilinear coordinates without additional geometric source terms. Since the requirements to satisfy the integral conservation laws are minimal, this work will facilitate the adoption of rigorous geometric effects in plasma fluid codes. Altogether, the novel form shown herein represents a curvilinear generalization of our earlier work\cite{Halpern2018}, in which we sought formulations of the fluid hierarchy that retain correct physical constraints when translated to computational implementations.

Resistive MHD exercises the full set of curvilinear operators, and therefore serves as a natural testbed for our work -- however, the formulation itself extends in straightforward manner to the entire fluid hierarchy. Using a set of simple but unconventional manipulations, and starting from the standard fluid equations, we arrive at a new formulation, Eqs.~\ref{eq_mass_anti}--\ref{eq_faraday}. These equations describe the evolution a set of generalized fluid quantities related to quadratic invariants of the system.  It is found that mass, angular momenta (for ignorable directions), and total energy are conserved in the new formulation. This result is, of course, unsurprising, since a well formulated fluid system should always conserve energy in the continuum.

There are, however, several interesting and less conventional results presented here. First, the conservation proofs only invoke the anti-symmetry of the first derivative, the solenoidal constraint $\nabla\cdot\vec{B}$, and the orthogonality of the scalar and cross products -- therefore, they possess natural discrete analogs. Second, the strategy allows us to express the fluid equations to be expressed in general curvilinear coordinates without the use of Christoffel symbols. We remark, in particular, the simplistic and practical expression obtained for the kinematic viscosity of a Newtonian fluid, the viscous heating, and the viscous fluxes. Taken together, the approach yields a simple but rigorous blueprint for achieving general curvilinear geometry with almost any numerical strategy.

A minimal implementation based on finite difference discretization is achieved using the \ALMA{} numerical engine. Then, the approach is validated using two representative problems. First, we use steady-state Shercliff flows to establish the overall correctness of the new formulation and its implementation. This included a test of a curvilinear mapping designed to increase resolution near steep shear layers.  We then consider the classic Orszag-Tang vortex problem to examine the numerical conservation properties in non-orthogonal curvilinear coordinates. Despite being evolved in a curvilinear coordinate system that is not adapted to the natural symmetries of the problem, the simulation remains qualitatively similar to its Cartesian counterpart. The novel formulation preserves the dominant flow structures while maintaining exact energy conservation. Overall, this comparison demonstrates and illustrates the governing role of underlying physical symmetries in the dynamics, and demonstrates the robustness of the approach in geometries that do not coincide with the natural coordinate directions.

\begin{acknowledgements}
FDH would like to thank P.~Putham (UKAEA) for developing the steady state flow case. This material is based upon work supported by the U.S. Department of Energy, Office of Science, Office of Fusion Energy Sciences, Theory Program, under Award DE-FG02-95ER54309. --Disclaimer: This report was prepared as an account of work sponsored by an agency of the United States Government. Neither the United States Government nor any agency thereof, nor any of their employees, makes any warranty, express or implied, or assumes any legal liability or responsibility for the accuracy, completeness, or usefulness of any information, apparatus, product, or process disclosed, or represents that its use would not infringe privately owned rights. Reference herein to any specific commercial product, process, or service by trade name, trademark, manufacturer, or otherwise, does not necessarily constitute or imply its endorsement, recommendation, or favoring by the United States Government or any agency thereof. The views and opinions of authors expressed herein do not necessarily state or reflect those of the United States Government or any agency thereof.
\end{acknowledgements}
\appendix

\section{Coordinate transformations and vector calculus}\label{AppA_transforms}
This is a summary of the standard curvilinear transforms and vector calculus operations\cite{Dhaeseleer1991} used in the manuscript:
\begin{align}
&\textbf{Coordinate transform:} &
\vec{x} &= \vec{x}(\xi^1,\xi^2,\xi^3), 
&
\partial_i &\equiv \frac{\partial}{\partial \xi^i}
\\&\textbf{Basis vectors} &
\vec{e}_i &= \partial_i \vec{x},
&
\vec{e}^i &= \nabla \xi^i
\\&\textbf{Jacobian} &
\jac &= \vec{e}_i \cdot (\vec{e}_j \times \vec{e}_k)
\\&\textbf{Metric factors} &
g_{ij} &= \vec{e}_i \cdot \vec{e}_j,
&
g^{ij} &= \vec{e}^i \cdot \vec{e}^j
\\&\textbf{Gradient} &
\nabla \phi &= \vec{e}^i \, \partial_i \phi
\\&\textbf{Divergence} &
\nabla \cdot \vec{A}
&= \frac{1}{\jac}\,\partial_i\!\left(\jac A^i\right)
\\&\textbf{Curl} &
(\nabla \times \vec{A})^i
&= \frac{1}{\jac}\,\epsilon_{ijk}\,\partial_j A_k
\\&\textbf{Dot product} &
\vec{A}\cdot\vec{B}
&= A_j B^j = g_{ij} A^i B^j
\\&\textbf{Cross product} &
(\vec{A}\times\vec{B})^i
&= \frac{1}{\jac}\,\epsilon_{ijk} A_j B_k
&(\vec{A}\times\vec{B})_i
&= {\jac}\,\epsilon^{ijk} A^j B^k
\end{align}Here $\epsilon_{ijk}$ and $\epsilon^{ijk}$ are commutator symbols for the lower and upper indices respectively. We usually denote the vector components as ``tangent'' ($A_i$) or ``reciprocal'' ($A^i$).

\section{Curvilinear form of the anti-symmetric representation}\label{AppB_curv}
It is possible to express the fluid equations in curvilinear coordinates using our previous anti-symmetry strategy.  This strategy was originally developed in Ref~\onlinecite{Halpern2021b}, and later successfully tested\cite{Koshkarov2022} using the classic MHD blast test\cite{Zachary1994}. This approach was sidelined due to (a) the difficulty of tracking and verifying Christoffel symbols in general curvilinear coordinates (b) the presence of non-linear quadratic source terms. Equation~\ref{eq_mom_anti} is replaced by
\begin{align}
    \partial_t \mm\,
    +& \frac{1}{2}\left(\gradd\cdot\vv + \vv\cdot\nabla\right)\mm +  \frac{\left(\gamma-1\right)}{\rr}\left(\UU\nabla\UU - \UU^2\vec{G}\right)
    + \jac\frac{\vec{B}}{\mu_0\rr}\times\curl\vec{B}
    + \frac{\jac}{\rr} \nabla \cdot \tau = 0, \label{eq_mom_anti3}
\end{align}with identical notation as Sec.~\ref{SecII_derivation}. The mass, internal energy, and induction equations evolved remain identical to Eqs.~\ref{eq_mass_anti}--\ref{eq_faraday}. However the divergence and advection operators acting on $\mm$ lead to quadratic, non-linear source terms multiplied by Christoffel symbols, \emph{i.e.} the components of Eq.~\ref{eq_mom_anti3} read
\begin{align}
    \partial_t \ubar{m}_i + \frac{1}{2}\left(\partial_j v^j + v^j \partial_j\right) \ubar{m}_i - \Gamma^k_{ij} v^j \ubar{m}_k + \ldots,
\end{align}where $\Gamma^k_{ij}$ are the Christoffel symbols of the second kind. It can be shown that energy is conserved by considering the reciprocal components,
\begin{align}
    \partial_t \ubar{m}^i + \frac{1}{2}\left(\partial_j v^j + v^j \partial_j\right) \ubar{m}^i + \Gamma^i_{jk} v^j \ubar{m}^k + \ldots
\end{align}We then compute $\partial_t |\mm|^2$ by multiplying each equation by $\ubar{m}^i$ and $\ubar{m}_i$ and adding the results. The non-linear source terms associated to the Christoffel symbols cancel exactly using the following manipulation
\begin{align}
    \ubar{m}^i \Gamma^k_{ij} v^j \ubar{m}_k = \ubar{m}_k \Gamma^k_{ij} \ubar{m}^i v^j = \ubar{m}_i \Gamma^{i}_{jk} v^j \ubar{m}^k.
\end{align}Energy conservation follows from anti-symmetry after integrating over volume, and adding the internal and magnetic energy. 
Note that while we use here the moniker ``anti-symmetric representation'', the force operator is no longer anti-symmetric as a result of the Christoffel symbols and other geometric source terms. This motivated the search for more general forms that would avoid geometric sources.

\section{Coordinate stretching and compression}\label{AppC_stretch}
Equation~\ref{eq_mom_anti3} has a useful form in the case of orthonormal coordinate systems, which was suggested by expressions appearing in the appendices of Ref.~\onlinecite{Satoh2004}. Defining $h_i = |\vec{e_i}|$, and replacing $v_i \rightarrow v_i / h_i$ and $v^i \rightarrow h_i v_i$ we find
\begin{align}
    \jac\left[\divp{\frac{\mm\mm}{\jac}}\right]_i = &
    \frac{1}{2}\rr\left( \partial_j\frac{v_j}{h_j}+\frac{v_j}{h_j}\partial_j\right)\ubar{m}_i + 
    \frac{1}{2}\ubar{m_i}\left( \partial_j\frac{v_j}{h_j}+\frac{v_j}{h_j}\partial_j\right)\rr \nonumber\\&
    + \left(\frac{\partial_j h_i}{h_j h_i}\right){\ubar{m}_i\ubar{m}_j}
    - \left(\frac{\partial_i h_j}{h_j h_i}\right){\ubar{m}_j\ubar{m}_j}.
\end{align}Here all vectors are expressed using their physical components. A suitable momentum equation follows by subtracting the mass transport $\frac{1}{2}\sim \ubar{m}_i\partial_t\rr$. This form is achieved thanks to the simplicity of the Christoffel symbols for a diagonal metric. The rest of the fluid system is expressed in straightforward manner using physical vector components for the velocity and magnetic field. The form above is particularly useful for mappings expressing pure 1D coordinate stretching or compressing, since the terms involving $\partial_j h_i$ and $\partial_i h_j$ vanish, and the viscosity tensor simplifies considerably compared to the standard expressions\cite{Holler2014}.

\section{Shock propagation speed}\label{AppD_Sod}
\begin{figure}
    \centering
    \includegraphics[width=0.48\linewidth]{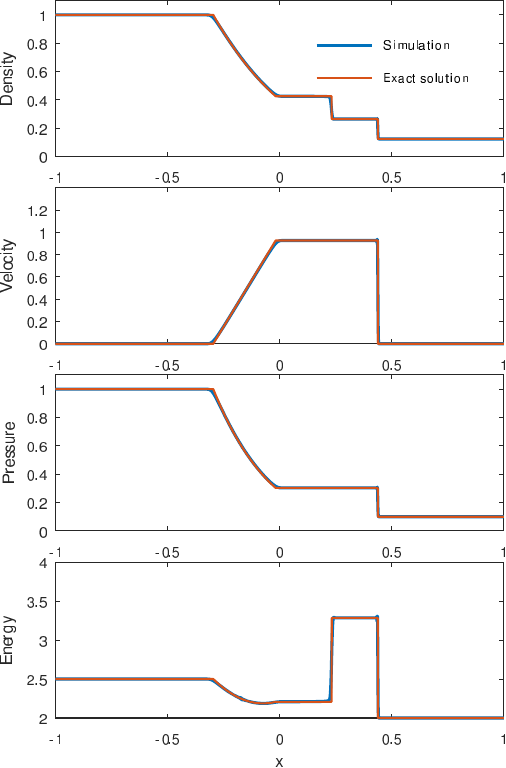}
    \includegraphics[width=0.48\linewidth]{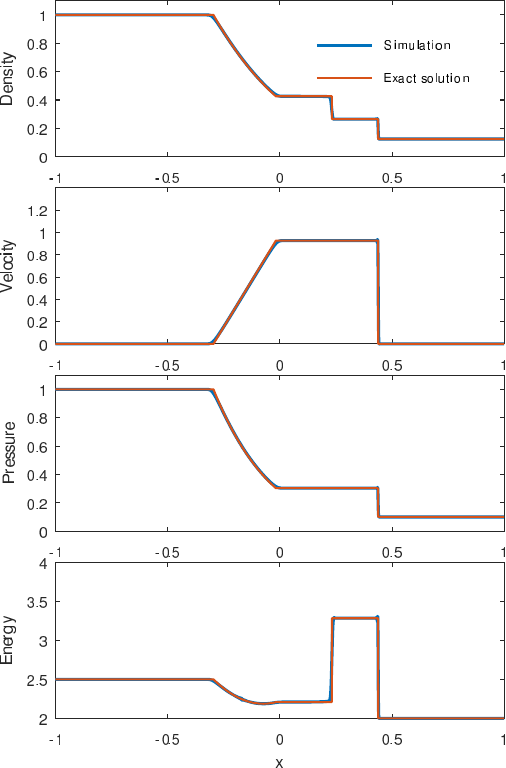}
    \caption{Fluid density, velocity, pressure, and energy (top to bottom) for simulations of the Sod shock tube test carried out using centered finite difference discretizations of either Eqs.~\ref{eq_mass_anti}--\ref{eq_pres_anti} (left) or the previous antisymmetry representation  (right).}
    \label{fig_Sod}
\end{figure}

As a basic consistency check, we consider the 1D Cartesian Sod shock tube problem\cite{Sod1978} using resolution $N_x=2000$, viscosity $\mu=10^{-4}$, and adiabatic index $\gamma=7/5$. Regularizing diffusion $D=\mu$ and heat-flux $\chi=2\mu$ are employed. The discretization is carried out with centered finite differences and 3rd order strong-stability-preserving Runge-Kutta method\cite{Shu1988}. We carry out the test using the present formulation (Eqs.~\ref{eq_mass_anti}--\ref{eq_pres_anti}), as well as the anti-symmetry formulation in Ref.~\onlinecite{Halpern2020} (left and right panels of Fig.~\ref{fig_Sod}, respectively. Both formulations reproduce the correct wave structure and shock propagation, reflecting consistency with the Rankine–Hugoniot conditions through the manifest symmetries of the force operator. Some small oscillations persist near discontinuities, as expected from non-monotone discretizations. Standard shock-capturing or filtering techniques (e.g., \onlinecite{Kennedy2008}) can be incorporated to enforce monotonicity if desired.

\bibliography{ALMA}

@article{Chacon2004,
title = {A non-staggered, conservative, ∇⋅B→=0, finite-volume scheme for 3D implicit extended magnetohydrodynamics in curvilinear geometries},
journal = {Computer Physics Communications},
volume = {163},
number = {3},
pages = {143-171},
year = {2004},
issn = {0010-4655},
doi = {10.1016/j.cpc.2004.08.005},
author = {L. Chac\'on}
}

@article{Cole2006,
    author = {Cole, A. and Fitzpatrick, R.},
    title = {Drift-magnetohydrodynamical model of error-field penetration in tokamak plasmas},
    journal = {Physics of Plasmas},
    volume = {13},
    number = {3},
    pages = {032503},
    year = {2006},
    month = {03},
    issn = {1070-664X},
    doi = {10.1063/1.2178167},
}

@book{Satoh2004,
  title={Atmospheric circulation dynamics and circulation models},
  author={Satoh, Masaki},
  year={2004},
  publisher={Springer Science \& Business Media}
}

@book{Dhaeseleer1991,
  title={Flux Coordinates and Magnetic Field Structure: A Guide to a Fundamental Tool of Plasma Theory},
  author={D'haeseleer, W.D. and Hitchon, W.N.G. and Callen, J.D. and Shohet, J.L.},
  doi={doi.org/10.1007/978-3-642-75595-8},
  isbn={9783642755958},
  lccn={90010333},
  series={Scientific Computation},
  year={1991},
  publisher={Springer Berlin Heidelberg}
}

@article{Zachary1994,
author = {Zachary, Andrew L. and Malagoli, Andrea and Colella, Phillip},
title = {A Higher-Order Godunov Method for Multidimensional Ideal Magnetohydrodynamics},
journal = {SIAM Journal on Scientific Computing},
volume = {15},
number = {2},
pages = {263-284},
year = {1994},
doi = {10.1137/0915019},
}

@article{Vinokur1974,
  title={Conservation equations of gasdynamics in curvilinear coordinate systems},
  author={Vinokur, Marcel},
  journal={Journal of Computational Physics},
  volume={14},
  number={2},
  pages={105--125},
  year={1974},
  doi = {10.1016/0021-9991(74)90008-4},
  publisher={Elsevier}
}

@article{Sjogreen2014,
title = {On high order finite-difference metric discretizations satisfying {GCL} on moving and deforming grids},
journal = {Journal of Computational Physics},
volume = {265},
pages = {211-220},
year = {2014},
issn = {0021-9991},
doi = {10.1016/j.jcp.2014.01.045},
author = {Björn Sjögreen and H.C. Yee and Marcel Vinokur}
}

@article{Porcelli1987,
    author = {Porcelli, Francesco},
    title = {Viscous resistive magnetic reconnection},
    journal = {The Physics of Fluids},
    volume = {30},
    number = {6},
    pages = {1734-1742},
    year = {1987},
    month = {06},
    issn = {0031-9171},
    doi = {10.1063/1.866240},
}

@article{Kennedy2008,
title = {Reduced aliasing formulations of the convective terms within the Navier–Stokes equations for a compressible fluid},
journal = {Journal of Computational Physics},
volume = {227},
number = {3},
pages = {1676-1700},
year = {2008},
issn = {0021-9991},
doi = {10.1016/j.jcp.2007.09.020},
author = {Christopher A. Kennedy and Andrea Gruber}
}

@article{Sod1978,
title = {A survey of several finite difference methods for systems of nonlinear hyperbolic conservation laws},
journal = {Journal of Computational Physics},
volume = {27},
number = {1},
pages = {1-31},
year = {1978},
issn = {0021-9991},
doi = {10.1016/0021-9991(78)90023-2},
author = {Gary A Sod}
}

@article{Powell1999,
title = {A Solution-Adaptive Upwind Scheme for Ideal Magnetohydrodynamics},
journal = {Journal of Computational Physics},
volume = {154},
number = {2},
pages = {284-309},
year = {1999},
issn = {0021-9991},
doi = {10.1006/jcph.1999.6299},
author = {Kenneth G. Powell and Philip L. Roe and Timur J. Linde and Tamas I. Gombosi and Darren L. {De Zeeuw}}
}

@misc{Koshkarov2022,
  author       = {Koshkarov, Oleksandr},
  note         = {Private communication},
  year         = {2022}
}

@article{Pfirsch1996,
   author = {Dieter Pfirsch and Dar\'io Correa-Restrepo},
   doi = {10.1088/0741-3335/38/1/003},
   issn = {07413335},
   issue = {1},
   journal = {Plasma Physics and Controlled Fusion},
   month = {1},
   pages = {71-101},
   title = {Collisional drift fluid equations and implications for drift waves},
   volume = {38},
   year = {1996},
}

@article{Oreilly2020,
title = {Energy conservative SBP discretizations of the acoustic wave equation in covariant form on staggered curvilinear grids},
journal = {Journal of Computational Physics},
volume = {411},
pages = {109386},
year = {2020},
issn = {0021-9991},
doi = {10.1016/j.jcp.2020.109386},
url = {https://arxiv.org/pdf/1907.01105},
author = {Ossian O'Reilly and N. Anders Petersson}
}

@article{Holler2014,
author = {H{\"{o}}ller, Harald and Koskela, Antti and Dorfi, Ernst and Benger, Werner},
doi = {10.1186/s40668-014-0002-6},
issn = {2197-7909},
journal = {Computational Astrophysics and Cosmology},
number = {1},
pages = {2},
title = {{Artificial viscosity in comoving curvilinear coordinates: towards a differential geometrically consistent implicit advection scheme}},
url = {https://doi.org/10.1186/s40668-014-0002-6},
volume = {1},
year = {2014}
}

@article{Smolentsev2015,
title = {An approach to verification and validation of MHD codes for fusion applications},
journal = {Fusion Engineering and Design},
volume = {100},
pages = {65-72},
year = {2015},
issn = {0920-3796},
doi = {10.1016/j.fusengdes.2014.04.049},
author = {S. Smolentsev and S. Badia and R. Bhattacharyay and L. Bühler and L. Chen and Q. Huang and H.-G. Jin and D. Krasnov and D.-W. Lee and E. Mas {de les Valls} and C. Mistrangelo and R. Munipalli and M.-J. Ni and D. Pashkevich and A. Patel and G. Pulugundla and P. Satyamurthy and A. Snegirev and V. Sviridov and P. Swain and T. Zhou and O. Zikanov}
}

@article{Shercliff1953,
title={Steady motion of conducting fluids in pipes under transverse magnetic fields},
volume={49},
DOI={10.1017/S0305004100028139},
number={1},
journal={Mathematical Proceedings of the Cambridge Philosophical Society},
author={Shercliff, J.~A.},
year={1953},
pages={136–144}
}

@article{Beer1995,
    author = {Beer, M. A. and Cowley, S. C. and Hammett, G. W.},
    title = "{Field‐aligned coordinates for nonlinear simulations of tokamak turbulence}",
    journal = {Physics of Plasmas},
    volume = {2},
    number = {7},
    pages = {2687-2700},
    year = {1995},
    month = {07},
    issn = {1070-664X},
    doi = {10.1063/1.871232},
    url = {https://doi.org/10.1063/1.871232}
}

@article{Boozer1981,
    author = {Boozer, Allen H.},
    title = {Plasma equilibrium with rational magnetic surfaces},
    journal = {The Physics of Fluids},
    volume = {24},
    number = {11},
    pages = {1999-2003},
    year = {1981},
    month = {11},
    issn = {0031-9171},
    doi = {10.1063/1.863297}
}

@article{Hamada1962,
doi = {10.1088/0029-5515/2/1-2/005},
year = {1962},
month = {jan},
publisher = {},
volume = {2},
number = {1-2},
pages = {23},
author = {Hamada, Shigeo},
title = {Hydromagnetic equilibria and their proper coordinates},
journal = {Nuclear Fusion}
}

@article{VonNeumann1932,
 ISSN = {0003486X},
 URL = {http://www.jstor.org/stable/1968537},
 author = {J. {von Neumann}},
 journal = {Annals of Mathematics},
 number = {3},
 pages = {587--642},
 publisher = {Annals of Mathematics},
 title = {Zur Operatorenmethode In Der Klassischen Mechanik},
 volume = {33},
 year = {1932}
}

@article {Koopman1931,
	author = {Koopman, B. O.},
	title = {Hamiltonian Systems and Transformation in Hilbert Space},
	journal = {Proceedings of the National Academy of Sciences},
	volume = {17},
	number = {5},
	pages = {315--318},
	year = {1931},
	doi = {10.1073/pnas.17.5.315},
	publisher = {National Academy of Sciences}
}

@article{Joseph2020,
  title = {Koopman--von Neumann approach to quantum simulation of nonlinear classical dynamics},
  author = {Joseph, Ilon},
  journal = {Phys. Rev. Research},
  volume = {2},
  issue = {4},
  pages = {043102},
  numpages = {17},
  year = {2020},
  month = {Oct},
  publisher = {American Physical Society},
  doi = {10.1103/PhysRevResearch.2.043102},
  url = {https://link.aps.org/doi/10.1103/PhysRevResearch.2.043102}
}

@article{Halpern2021,
title = {Simulations of plasmas and fluids using anti-symmetric models},
journal = {Journal of Computational Physics},
volume = {445},
pages = {110631},
year = {2021},
issn = {0021-9991},
doi = {10.1016/j.jcp.2021.110631},
url = {https://www.sciencedirect.com/science/article/pii/S002199912100526X},
author = {Federico D. Halpern and Igor Sfiligoi and Mark Kostuk and Ryan Stefan and Ronald E. Waltz}
}

@article{Halpern2020,
author = {Halpern,Federico D. },
title = {Anti-symmetric representation of the extended magnetohydrodynamic equations},
journal = {Physics of Plasmas},
volume = {27},
number = {4},
pages = {042303},
year = {2020},
doi = {10.1063/5.0002345}
}

@article{Shu1988,
author = {Shu, Chi-Wang},
title = {Total-Variation-Diminishing Time Discretizations},
journal = {SIAM Journal on Scientific and Statistical Computing},
volume = {9},
number = {6},
pages = {1073-1084},
year = {1988},
doi = {10.1137/0909073},
URL = {https://doi.org/10.1137/0909073},
}

@article{Orszag1979,
title={Small-scale structure of two-dimensional magnetohydrodynamic turbulence},
volume={90},
DOI={10.1017/S002211207900210X},
number={1},
journal={Journal of Fluid Mechanics},
publisher={Cambridge University Press},
author={Orszag, Steven A. and Tang, Cha-Mei},
year={1979},
pages={129--143}}

@article{Jardin2012,
  author={S C Jardin and N Ferraro and J Breslau and J Chen},
  title={Multiple timescale calculations of sawteeth and other global macroscopic dynamics of tokamak plasmas},
  journal={Computational Science \& Discovery},
  volume={5},
  number={1},
  pages={014002},
  url={http://stacks.iop.org/1749-4699/5/i=1/a=014002},
  year={2012},
}

@article{Sovinec2004,
title = "Nonlinear magnetohydrodynamics simulation using high-order finite elements",
journal = "Journal of Computational Physics",
volume = "195",
number = "1",
pages = "355 - 386",
year = "2004",
issn = "0021-9991",
doi = "https://doi.org/10.1016/j.jcp.2003.10.004",
url = "http://www.sciencedirect.com/science/article/pii/S0021999103005369",
author = "C.R. Sovinec and A.H. Glasser and T.A. Gianakon and D.C. Barnes and R.A. Nebel and S.E. Kruger and D.D. Schnack and S.J. Plimpton and A. Tarditi and M.S. Chu"
}

@article{Czarny2008,
title = "B\'{e}zier surfaces and finite elements for {MHD} simulations",
journal = "Journal of Computational Physics",
volume = "227",
number = "16",
pages = "7423 - 7445",
year = "2008",
issn = "0021-9991",
doi = "10.1016/j.jcp.2008.04.001",
author = "Olivier Czarny and Guido Huysmans"
}

@article{Butcher1964,
	title={Implicit runge-kutta processes},
	author={Butcher, John C},
	journal={Mathematics of Computation},
	volume={18},
	number={85},
	pages={50--64},
	year={1964},
	doi =  {10.1090/S0025-5718-1964-0159424-9}
}

@article{Strauss1976,
	author = {H. R. Strauss},
	title = {Nonlinear, threeâdimensional magnetohydrodynamics of noncircular tokamaks},
	journal = {The Physics of Fluids},
	volume = {19},
	number = {1},
	pages = {134-140},
	year = {1976},
	doi = {10.1063/1.861310},
	
}

@article{Lust1959,
author = {L\"{u}st, Reimar},
title = {Über die Ausbreitung von Wellen in einem Plasma},
journal = {Fortschritte der Physik},
volume = {7},
number = {9},
pages = {503-558},
year = {1959},
doi = {10.1002/prop.19590070902},
}

@article{Halpern2018,
author = {Halpern,Federico D.  and Waltz,Ronald E. },
title = {Anti-symmetric plasma moment equations with conservative discrete counterparts},
journal = {Physics of Plasmas},
volume = {25},
number = {6},
pages = {060703},
year = {2018},
doi = {10.1063/1.5038110}
}

@article{Halpern2021b,
author = {Halpern,Federico D. and Bernard, Tess N. and Waltz, Ronald E. },
title = {Curvilinear formulation of anti-symmetric plasmas models},
journal = {Bulletin of the American Physical Society},
volume = {66},
number = {13},
pages = {GP11.00044},
year = {2021},
address = {Pittsburg, PA},
publisher = {Soc.},
url = {https://meetings.aps.org/Meeting/DPP21/Session/GP11.44}
}

@ARTICLE{Arakawa1966,
  author = {Akio Arakawa},
  title = {Computational design for long-term numerical integration of the equations
	of fluid motion: Two-dimensional incompressible flow. Part I},
  journal = {Journal of Computational Physics},
  year = {1966},
  volume = {1},
  pages = {119 - 143},
  number = {1},
  doi = {10.1016/0021-9991(66)90015-5},
  issn = {0021-9991},
  url = {http://www.sciencedirect.com/science/article/pii/0021999166900155}
}

@book{Biskamp2000,
  title={Magnetic Reconnection in Plasmas},
  author={Biskamp, D. and Haines, M.G. and Hopcraft, K.I. and Hutchinson, I.H. and Schindler, K. and Surko, C.M.},
  isbn={9780521582889},
  lccn={99087680},
  series={Cambridge Monographs on Plasma Physics},
  year={2000},
  publisher={Cambridge University Press}
}

@BOOK{Braginskii1965,
  title = {Transport processes in a plasma},
  publisher = {Consultants Bureau},
  year = {1965},
  editor = {M. A. Leontovich},
  author = {S. I. Braginskii},
  volume = {1},
  pages = {205},
  series = {Reviews of Plasma Physics},
  address = {New York}
}

@ARTICLE{Connor1978,
  author = {Connor, J. W. and Hastie, R. J. and Taylor, J. B.},
  title = {Shear, Periodicity, and Plasma Ballooning Modes},
  journal = {Phys. Rev. Lett.},
  year = {1978},
  volume = {40},
  pages = {396--399},
  month = {Feb},
  doi = {10.1103/PhysRevLett.40.396},
  issue = {6},
  publisher = {American Physical Society},
  url = {http://link.aps.org/doi/10.1103/PhysRevLett.40.396}
}

@ARTICLE{Dudson2009,
  author = {B.D. Dudson and M.V. Umansky and X.Q. Xu and P.B. Snyder and H.R.
	Wilson},
  title = {{BOUT++:} A framework for parallel plasma fluid simulations },
  journal = {Computer Physics Communications },
  year = {2009},
  volume = {180},
  pages = {1467 - 1480},
  number = {9},
  doi = {10.1016/j.cpc.2009.03.008},
  issn = {0010-4655},
  url = {http://www.sciencedirect.com/science/article/pii/S0010465509001040}
}

@article{Fryxell2000,
	doi = {10.1086/317361},
	url = {https://doi.org/10.1086%2F317361},
	year = 2000,
	month = {nov},
	publisher = {{IOP} Publishing},
	volume = {131},
	number = {1},
	pages = {273--334},
	author = {B. Fryxell and K. Olson and P. Ricker and F. X. Timmes and M. Zingale and D. Q. Lamb and P. MacNeice and R. Rosner and J. W. Truran and H. Tufo},
	title = {{FLASH}: An Adaptive Mesh Hydrodynamics Code for Modeling Astrophysical Thermonuclear Flashes},
	journal = {The Astrophysical Journal Supplement Series}
}

@ARTICLE{Morinishi2004,
  author = {Youhei Morinishi and Oleg V. Vasilyev and Takeshi Ogi},
  title = {Fully conservative finite difference scheme in cylindrical coordinates
	for incompressible flow simulations },
  journal = {Journal of Computational Physics },
  year = {2004},
  volume = {197},
  pages = {686 - 710},
  number = {2},
  doi = {http://dx.doi.org/10.1016/j.jcp.2003.12.015},
  issn = {0021-9991},
  url = {http://www.sciencedirect.com/science/article/pii/S0021999103006594}
}

@ARTICLE{Scott1997,
  author = {B Scott},
  title = {Three-dimensional computation of drift Alfvén turbulence},
  journal = {Plasma Physics and Controlled Fusion},
  year = {1997},
  volume = {39},
  pages = {1635},
  number = {10},
  url = {http://stacks.iop.org/0741-3335/39/i=10/a=010}
}

@book{Shashkov1996,
author = {Mikhail J. Shashkov},
title = {Conservative finite-difference methods on general grids},
date = {1996},
year = {1996},
editor = {Stanley Steinberg},
publisher = {CRC Press},
location = {New York},
isbn = {9780849373756},
}

@article{felker2018,
title = "A fourth-order accurate finite volume method for ideal MHD via upwind constrained transport",
journal = "Journal of Computational Physics",
volume = "375",
pages = "1365 - 1400",
year = "2018",
issn = "0021-9991",
doi = "10.1016/j.jcp.2018.08.025",
url = "http://www.sciencedirect.com/science/article/pii/S0021999118305485",
author = "Kyle Gerard Felker and James M. Stone"
}

@article{Chew1956,
    author = {Chew, G. F. and Goldberger, M. L. and Low, F. E.},
    title = {The Boltzmann equation an d the one-fluid hydromagnetic equations in the absence of particle collisions},
    journal = {Proceedings of the Royal Society of London. A. Mathematical and Physical Sciences},
    volume = {236},
    number = {1204},
    pages = {112-118},
    year = {1956},
    month = {07},
    doi = {10.1098/rspa.1956.0116}
}

@article{Craxton2015,
    author = {Craxton, R. S. and Anderson, K. S. and Boehly, T. R. and Goncharov, V. N. and Harding, D. R. and Knauer, J. P. and McCrory, R. L. and McKenty, P. W. and Meyerhofer, D. D. and Myatt, J. F. and Schmitt, A. J. and Sethian, J. D. and Short, R. W. and Skupsky, S. and Theobald, W. and Kruer, W. L. and Tanaka, K. and Betti, R. and Collins, T. J. B. and Delettrez, J. A. and Hu, S. X. and Marozas, J. A. and Maximov, A. V. and Michel, D. T. and Radha, P. B. and Regan, S. P. and Sangster, T. C. and Seka, W. and Solodov, A. A. and Soures, J. M. and Stoeckl, C. and Zuegel, J. D.},
    title = {Direct-drive inertial confinement fusion: A review},
    journal = {Physics of Plasmas},
    volume = {22},
    number = {11},
    pages = {110501},
    year = {2015},
    month = {11},
    issn = {1070-664X},
    doi = {10.1063/1.4934714}
}

@ARTICLE{Zeiler1998,
  author = {A. Zeiler and D. Biskamp and J. F. Drake and B. N. Rogers},
  title = {Transition from resistive ballooning to $\eta_i$ driven turbulence
	in tokamaks},
  journal = {Physics of Plasmas},
  year = {1998},
  volume = {5},
  pages = {2654-2663},
  number = {7},
  doi = {10.1063/1.872953},
  publisher = {AIP},
  url = {http://link.aip.org/link/?PHP/5/2654/1}
}

@ARTICLE{Zeiler1997,
  author = {Zeiler, A. and Drake, J. F. and Rogers, B.},
  title = {Nonlinear reduced {Braginskii} equations with ion thermal dynamics
	in toroidal plasma},
  journal = {Physics of Plasmas},
  year = {1997},
  volume = {4},
  pages = {2134-2138},
  number = {6},
  doi = {10.1063/1.872368}
}

@article{Zweibel2009,
author = {Zweibel, Ellen G. and Yamada, Masaaki},
title = {Magnetic Reconnection in Astrophysical and Laboratory Plasmas},
journal = {Annual Review of Astronomy and Astrophysics},
volume = {47},
number = {1},
pages = {291-332},
year = {2009},
doi = {10.1146/annurev-astro-082708-101726}
}

@article{Lindl1995,
    author = {Lindl, John},
    title = {Development of the indirect‐drive approach to inertial confinement fusion and the target physics basis for ignition and gain},
    journal = {Physics of Plasmas},
    volume = {2},
    number = {11},
    pages = {3933-4024},
    year = {1995},
    month = {11},
    issn = {1070-664X},
    doi = {10.1063/1.871025}
}
\end{document}